\documentclass[trackchanges,twocolumn]{aastex7} 

\usepackage{amsmath}
\usepackage{dblfloatfix}
\usepackage{float} 
\usepackage{gensymb}
\usepackage{wrapfig} 
\usepackage[normalem]{ulem}  

\begin{document}

\title{XRISM finds the Changing-Look AGN NGC 1365 in an extended low state: \\ A dense, highly ionized outflow obscures the central source}

\author[orcid=0000-0003-0931-0868]{Fatima Zaidouni}
\affiliation{MIT Kavli Institute for Astrophysics and Space Research, Massachusetts Institute of Technology, Cambridge, MA 02139, USA}
\email[show]{fzaid@mit.edu}  

\author[0000-0003-0172-0854]{Erin Kara}
\affiliation{MIT Kavli Institute for Astrophysics and Space Research, Massachusetts Institute of Technology, Cambridge, MA 02139, USA}
\email{ekara@mit.edu}

\author[orcid=0000-0003-4511-8427]{Peter Kosec}
\affiliation{Center for Astrophysics $|$ Harvard~\&~Smithsonian, 60 Garden St., Cambridge, MA 02138, USA}
\email{peter.kosec@cfa.harvard.edu}

\author[0000-0001-9735-4873]{Ehud Behar} %
\affiliation{Department of Physics, Technion, Technion City, Haifa 3200003,
Israel}
\email{behar@mit.edu}

\author[orcid=0000-0002-7962-5446]{Richard Mushotzky}
\affiliation{Astronomy Dept and JSI Joint Space Sciences Institute (JSI), University of Maryland , College Park MD 20740 USA}
\email{fake}

\author[0000-0002-7998-9581]{Michael Koss}
\affiliation{Eureka Scientific, Inc., 2452 Delmer Street, Suite 100, Oakland, CA 94602-3017, USA}
\email{Mike.Koss@eurekasci.com} 

\author[orcid=0000-0002-7292-6852,gname='Anna',sname='Jur\'{a}\v{n}ov\'{a}']{A.~Jur\'{a}\v{n}ov\'{a}}
\affiliation{MIT Kavli Institute for Astrophysics and Space Research, Massachusetts Institute of Technology, Cambridge, MA 02139, USA}
\email{ajuran@mit.edu}

\author[orcid=0000-0002-0273-218X]{Elias Kammoun}
\affiliation{Cahill Center for Astronomy \& Astrophysics, California Institute of Technology, Pasadena, CA 91125, USA}
\email{ekammoun@caltech.edu} 


\author[orcid=0000-0003-2663-1954]{Laura~W.~Brenneman}
\affiliation{Center for Astrophysics $|$ Harvard~\&~Smithsonian, 60 Garden St., Cambridge, MA 02138, USA}
\email{lbrenneman@cfa.harvard.edu}

\author[orcid=0000-0002-0568-6000]{Joheen Chakraborty}
\affiliation{MIT Kavli Institute for Astrophysics and Space Research, Massachusetts Institute of Technology, Cambridge, MA 02139, USA}
\email{joheen@mit.edu}

\author[0000-0002-5352-7178]{Ken Ebisawa} %
\affiliation{Institute of Space and Astronautical Science (ISAS), 
Japan Aerospace Exploration Agency (JAXA), Kanagawa 252-5210, Japan}
\email{ebisawa.ken@jaxa.jp}

\author[orcid=0000-0003-3894-5889]{Megan E. Eckart}
\affiliation{Lawrence Livermore National Laboratory, Livermore, CA 94550, USA}
\email{eckart2@llnl.gov}

\author[0000-0002-9378-4072]{Andrew C. Fabian} %
\affiliation{Institute of Astronomy, Cambridge University, Madingley Rd., Cambridge, CB3 0HA, UK}
\email{acf@ast.cam.ac.uk}

\author[0000-0002-0921-8837]{Yasushi Fukazawa} 
\affiliation{Department of Physics,
Hiroshima University, Hiroshima 739-8526, Japan}
\email{fukazawa@astro.hiroshima-u.ac.jp}

\author[0000-0003-3828-2448]{Javier A. García} 
\affiliation{NASA / Goddard Space Flight Center, Greenbelt, MD 20771, USA}
\affiliation{Cahill Center for Astronomy \& Astrophysics, California Institute of Technology, Pasadena, CA 91125, USA}
\email{javier@caltech.edu}

\author[0000-0001-9911-7038]{Liyi Gu} %
\affiliation{SRON Space Research Organisation Netherlands, Leiden, The Netherlands} 
\email{L.Gu@sron.nl}

\author[orcid=0000-0003-4127-0739]{Megan Masterson}
\affiliation{MIT Kavli Institute for Astrophysics and Space Research, Massachusetts Institute of Technology, Cambridge, MA 02139, USA}
\email{mmasters@mit.edu}

\author[0000-0002-5701-0811]{Shoji Ogawa}
\affiliation{Institute of Space and Astronautical Science (ISAS), Japan
Aerospace Exploration Agency (JAXA), Kanagawa 252-5210, Japan}
\email{sogawa@ac.jaxa.jp}

\author[orcid=0000-0002-6054-3432]{Takashi Okajima}
\affiliation{NASA Goddard Space Flight Center, Greenbelt, MD 20771, USA}
\email{takashi.okajima@nasa.gov}

\author[orcid=0000-0002-8108-9179]{St\'ephane Paltani}
\affiliation{Department of Astronomy, University of Geneva, 1290
Versoix, Switzerland}
\email{Stephane.Paltani@unige.ch}

\author[0000-0002-5359-9497]{Daniele Rogantini}
\affiliation{Department of Astronomy and Astrophysics, University of Chicago, 5640 S Ellis Ave, Chicago, IL 60637, USA}
\email{danieler@uchicago.edu}

\author[orcid=0000-0003-1780-5481]{Yuichi Terashima}
\affiliation{Department of Physics, Ehime University, Ehime 790-8577, Japan}
\email{terashima.yuichi.mc@ehime-u.ac.jp}

\author[0000-0003-2063-381X]{Brian J. Williams} %
\affiliation{NASA / Goddard Space Flight Center, Greenbelt, MD 20771, USA}
\email{brian.j.williams@nasa.gov}

\author[0000-0002-9754-3081]{Satoshi Yamada}
\affiliation{Frontier Research Institute for Interdisciplinary Sciences, Tohoku University, Sendai 980-8578, Japan}
\email{satoshi.yamada@astr.tohoku.ac.jp}

\begin{abstract}

We present the first XRISM/Resolve observations of the active galactic nucleus, NGC 1365, obtained in 2024 February and July. NGC 1365 is known for rapid transitions between Compton-thick and Compton-thin states, along with strong absorption from a highly ionized wind. During our observations, the source is found in a persistent low-flux state, characterized by a decrease in hard-X-ray luminosity and significant line-of-sight obscuration. In this state, XRISM/Resolve reveals clear Fe\,\textsc{xxv} and Fe\,\textsc{xxvi} absorption lines together with, for the first time in this source, corresponding emission lines. These features may arise either from reemission from a photoionized wind (P Cygni profile) or from collisionally ionized gas associated with outflow-driven shocks in the interstellar medium. We estimate the wind launch radius to be approximately $10^{16}~\mathrm{cm}$ ($\sim 10^4 R_{\mathrm{g}}$), consistent with the location of the X-ray broad-line region. We also resolve a broadened Fe K$\alpha$ line by $\sigma \sim 1300$ km s$^{-1}$ placing it at similar scales to the wind, consistent with radii inferred from disk-broadening models and the variability of the Fe K$\alpha$ broad line. The similarity of the Fe K$\alpha$ profile to the H$\beta$ wing and broad Pa$\alpha$ width indicates that the X-ray-emitting region is likely cospatial with the optical/IR broad-line region and originates from the same gas.

\end{abstract}


\let\clearpage\relax
\section{Introduction} \label{sec:intro}

\noindent Active Galactic Nuclei (AGN) are among the most energetic phenomena in the universe. Supermassive black holes (SMBHs) reside at the centers of most (if not all) galaxies, and at some point in their evolution, experienced epochs of intense accretion of gas. The accretion process simultaneously grows the black hole, while also releasing energy in the form of jets and outflows including ultra-fast outflows (UFOs), obscurers, and warm absorbers (e.g., \citealt{2021NatAs...5...13L}) and in the form of radiation. Wide-angle, high-velocity outflows are strong candidates for AGN feedback, a process through which SMBHs can significantly impact the evolution of their host galaxies, potentially regulating star formation rates on large kpc scales (\citealt{2012ARA&A..50..455F}). High-resolution X-ray spectroscopy provides a crucial means of detecting and characterizing outflows near their launching point, from milli-parsec scales and smaller, that are otherwise spatially unresolved. The launch of the X-ray Imaging and Spectroscopy Mission (XRISM) in September 2023 marks a significant advance in high-resolution studies of AGN outflows, providing new insights into the structure, dynamics, and impact of these winds (e.g., \citealt{2022IJMPD..3130001T}).

NGC 1365 ($z = 0.00547$; average redshift from NED\footnote{The NASA/IPAC Extragalactic Database: \url{http://ned.ipac.caltech.edu}}) was one of the first AGN targets observed with XRISM. It is a barred spiral galaxy in the Fornax galaxy cluster \citep{1981rsac.book.....S}, hosting a SMBH with an estimated mass of $\log (M_\mathrm{BH}/M_\odot) =  6.65 \pm 0.09$ (e.g, \citealt{2017MNRAS.468L..97O}; \citealt{2022ApJS..261....2K}). The nucleus of NGC 1365 hosts a so-called `Changing-Look AGN' that shows dramatic transitions between Compton-thin ($N_{\mathrm{H}} < 10^{24}$ cm$^{-2}$) and Compton-thick ($N_{\mathrm{H}} > 10^{24}$ cm$^{-2}$) states (e.g, \citealt{2005ApJ...623L..93R}). Historical variability in the line-of-sight column density ranges from a few $\times$ $10^{22}$ cm$^{-2}$ to $10^{24}$ cm$^{-2}$ (e.g., \citealt{2025A&A...693A..35J}; \citealt{2009MNRAS.393L...1R}) on short timescales up to a few tens of ks, which have been attributed to comet-like clouds in the broad-line region moving through our line of sight (BLR; \citealt{2010A&A...517A..47M}).

Outflowing, highly ionized gas has also been detected in this source. \citet{2005ApJ...630L.129R} first reported the detection of blueshifted absorption lines from Fe\,\textsc{xxv} and Fe\,\textsc{xxvi} in XMM-Newton CCD observations taken in 2004, providing early evidence for an ionized obscurer. Subsequent studies have shown that this obscurer persisted for over two decades, generally with ionization parameter $\xi>1000$~erg~cm~s$^{-1}$, outflow velocities in the range $\sim$2000--5000 km s$^{-1}$, and column densities between $10^{22}$ and $5 \times 10^{23}$ cm$^{-2}$ (e.g, \citealt{2005ApJ...630L.129R}; \citealt{2013MNRAS.429.2662B}). This outflow variability typically occurs within timescales as short as weeks, suggesting that the absorbing material is located relatively close to the central X-ray source, at distances comparable to the outer BLR (\citealt{2013MNRAS.429.2662B}).

In addition to its active nucleus, NGC 1365 is also a starburst galaxy hosting a circumnuclear starburst ring at around a $\sim1.8$ kpc distance away from the black hole (e.g, \citealt{2023ApJ...944L..15S}), whose emission dominates the soft X-ray emission ($\lesssim$2 keV; e.g.,\citealt{2009ApJ...694..718W}; \citealt{2015MNRAS.453.2558N}; \citealt{2009A&A...505..589G}). As a result, the X-ray spectrum includes contributions from both AGN-related photoionized emission and starburst-driven collisional processes. This makes NGC 1365 a particularly valuable system for investigating the interplay between black hole accretion and star formation in galaxy nuclei, including the role of feedback processes.

In this paper, we present the first observations of NGC 1365 with XRISM. XRISM is equipped with two instruments: XRISM/XTend, a CCD that provides a wide field of view and broad-band X-ray coverage over 0.3--12 keV, and XRISM/Resolve, the mission's prime instrument which will be the focus of this paper. Resolve is a high-resolution X-ray microcalorimeter spectrometer with an exceptional energy resolution of $\sim$5 eV at 6 keV, providing an unprecedented view of the X-ray spectral features of NGC 1365 (\citealt{10.1117/1.JATIS.11.4.042023}, Kelley et al. 2025 (accepted)). In Section~\ref{sec:data}, we describe the XRISM observations and data reduction process. Section~\ref{sec:modeling} presents the spectral modeling and key results, and in Section~\ref{sec:discussion}, we discuss their physical implications in the broader context of this AGN’s history. A summary of our findings is provided in Section~5.

In addition to the new XRISM Resolve observations, we use complementary data from multiple X-ray missions to provide context for our XRISM observation. Specifically, we use Chandra/HETG to compare the Fe K$\alpha$ complex at high resolution, long-term monitoring with Swift/BAT+XRT and NICER to track the light curve and source state, and contemporaneous XMM-Newton/EPIC-pn data to extend the broadband spectral coverage and compare to archival observations.

\section{Observations and Data Reduction} \label{sec:data}

\subsection{XRISM/Resolve Data Reduction}\label{subsec:xrism}

\begin{table}[t!]
\centering
\begin{tabular}{ccc}
\hline\hline
Obs. ID & Start Time & Exposure \\
\hline
Obs1: 300075010 & 20 Feb 2024 - 12:23:04 & 198 ks \\
Obs2: 300075020 & 03 Jul 2024 - 11:48:04 & 259 ks \\
\hline
\end{tabular}
\caption{Details of the two XRISM Resolve observations of NGC 1365.}
\label{tab:obs} 
\end{table}

\noindent The  XRISM/Resolve observations were taken in February 2024 (Obs1) and July 2024 (Obs2) during the Performance Verification phase (see Table~\ref{tab:obs} for details). We processed the clean event files using the XRISM software integrated in HEASoft v6.34 and used CALDB version 20241115 (XRISM CalDB 10). The standard reduction procedures were followed as outlined in the XRISM Quick Start Guide (v2.3) and the XRISM ABC Data Reduction Guide (v1.0)\footnote{\url{https://heasarc.gsfc.nasa.gov/docs/xrism/analysis/}}.

Spectra were extracted from the full Resolve array, excluding the calibration pixel 12 and pixel 27, which exhibits abnormal gain behavior. We used only High-resolution Primary (Hp) grade events, corresponding to X-ray events that do not overlap in time (within $\pm70.72$~ms) with other events in the same pixel \citep{10.1117/1.JATIS.4.1.011217}. These account for approximately 60\% of the total events in both observations. Approximately 38\% of events are classified as Low-resolution Secondary (LS). Low-resolution events are spurious events caused by a known issue with the onboard pulse-shape processing algorithm. Following current recommendations for weak sources, we exclude all low-resolution events from the event file used to generate the response matrix and all further analysis.

As these observations were taken near solar maximum, we needed to filter the observations (in time) for solar X-ray scattering events by identifying peaks in the simultaneous X-ray flux measured by the Geostationary Operational Environmental Satellite (GOES)\footnote{\url{https://www.swpc.noaa.gov/products/goes-x-ray-flux}}. In particular, we corrected for the strongest solar flares (class~X: Flux $> 10^{-4}$ W~m$^{-2}$), which removed approximately 150~s of data from Obs1, eliminating a spurious Cr line near 5.4~keV. No class~X flares were detected during Obs2. Filtering for less intense flares (class~M), or applying a threshold cut above the nominal background level in the simultaneous Xtend light curve, had no effect on the resulting spectra.

The response matrix files (RMFs) and ancillary response files (ARFs) were generated using the \texttt{rslmkrmf} and \texttt{xaarfgen} tools provided in HEASoft. For the final analysis, we adopted an ``extra-large'' RMF, which includes all components of the instrument line-spread function. The non–X-ray background (NXB) spectrum was generated using \texttt{rslnxbgen} and modeled using the public NXB spectral templates developed by the XRISM team\footnote{\url{https://heasarc.gsfc.nasa.gov/docs/xrism/analysis/nxb/}}. The Resolve observation was taken with the gate valve closed, resulting in no significant source signal below $\sim$1.7~keV, and the spectra generated with default settings, do not have good statistics above 10~keV. Accordingly, we use the 2–10~keV energy range throughout our analysis.

\subsection{XMM Newton Data Reduction}
\noindent We processed the EPIC-pn data from the XMM-Newton observation (obsid: 0924160101) that were taken concurrently with the Feb 2024 XRISM observation (Obs1) using the Science Analysis System (SAS v20.0.0) and the latest calibration files. The RGS data will be part of future work focusing on broadband modeling. The observation was conducted in Large Window mode, and data reduction followed standard procedures outlined in the XMM-Newton data analysis threads. The source spectrum was extracted from a circular region with a 35\arcsec\ radius centered on the source. The background was taken from a rectangular box (253\arcsec\,$\times$\,77.3\arcsec), which avoids contamination from the instrumental Cu line and ensures good background statistics. Flaring was filtered using a count rate threshold of 0.4\,counts\,s$^{-1}$, based on the single-event, high-energy (10--12\,keV) pn background light curve. We excluded soft proton background flares that were present at the beginning and end of the observation, as well as during a few short intervals in the middle. The resulting exposure time was 96\,ks. Response matrices were generated with  \texttt{rmfgen} and \texttt{arfgen}. The EPIC-pn spectra were binned to have at least 25 counts for each background-subtracted spectral channel and without oversampling the intrinsic energy resolution by a factor larger than 3. The final reduced EPIC-pn spectrum is presented in Fig.~\ref{fig:XMM} along with three other XMM observations from the public archive (obsids: 0205590301, 0505140201, 0692840401 in chronological order), which were extracted in a similar manner.

Figure~\ref{fig:Fig_lc_XMM.pdf} shows the EPIC-pn light curves extracted in the 2–4 keV and 4–10 keV bands, following standard XMM-Newton analysis threads and applying all relevant corrections using the \texttt{epiclccorr} task.

\subsection{NICER Data Reduction} 
\noindent The NICER X-ray Timing Instrument \citep{Gendreau16} aboard the International Space Station subsequently observed NGC 1365 over 52~ks across 85 observations (ObsIDs 6204130101-6204130119 and 7204130101-7204130166) between Jan. 30, 2024-Oct. 29, 2024. We followed the time-resolved spectroscopy approach for reliable estimation of source light curves outlined in Section 2.1 of \cite{Chakraborty24}.
Spectral fitting and background estimation was performed with the \texttt{SCORPEON}\footnote{\href{https://heasarc.gsfc.nasa.gov/docs/nicer/analysis_threads/scorpeon-overview/}{https://heasarc.gsfc.nasa.gov/docs/nicer/\\analysis\_threads/scorpeon-overview}} model over a broadband energy range (0.25--10 keV) for data taken in orbit night, and a slightly restricted range (0.38--10 keV) during orbit day. \texttt{SCORPEON} is a semi-empirical, physically motivated background model which explicitly includes components for the cosmic X-ray background as well as non X-ray noise events (e.g. precipitating electrons and cosmic rays) and can be fit along with the source to allow joint estimation of uncertainties. We grouped our spectra with the optimal binning scheme of \cite{Kaastra16}, i.e. \texttt{grouptype=optmin} with \texttt{groupscale=10} in the \texttt{ftgrouppha} command, and performed all spectral fitting with the Cash statistic \citep{Cash1979}. The light curve thus generated is presented in Fig.~\ref{fig:lc}.

\subsection{Chandra/HETG Data Reduction} 
\noindent We obtained 13 Chandra HETG \citep{2005PASP..117.1144C} observations of NGC 1365 from the Chandra archive taken from 2006 to 2021 (obsids: 24737, 24787, 24789, 24790, 24791, 24792, 24793, 24794, 24795, 25015, 25024, 13920, 13921) and reprocessed the data using CIAO \citep[\texttt{v14.12}][]{2006SPIE.6270E..1VF} and CALDB (\texttt{v4.9.2.1}). We followed standard data reduction techniques for HETG data, but reduced the width of the mask in the grating arms to 18 pixels, roughly half of the default value. This choice decreases the overlap between the High Energy Grating (HEG) and Medium Energy Grating (MEG), thereby allowing us to extend our HEG analysis to higher energies. We extracted the first order spectra for all observations and combined all of the positive and negative order HEG data with the \texttt{\detokenize{combine_grating_spectra}} tool. Fig. \ref{fig:chandra} shows a comparison of the combined Chandra HETG spectra (450~ks total) superimposed on the combined XRISM spectrum.

\subsection{XRISM Observations in Context} \label{sec:context}

\begin{figure}[t!]
    \centering
    \includegraphics[width=\columnwidth]{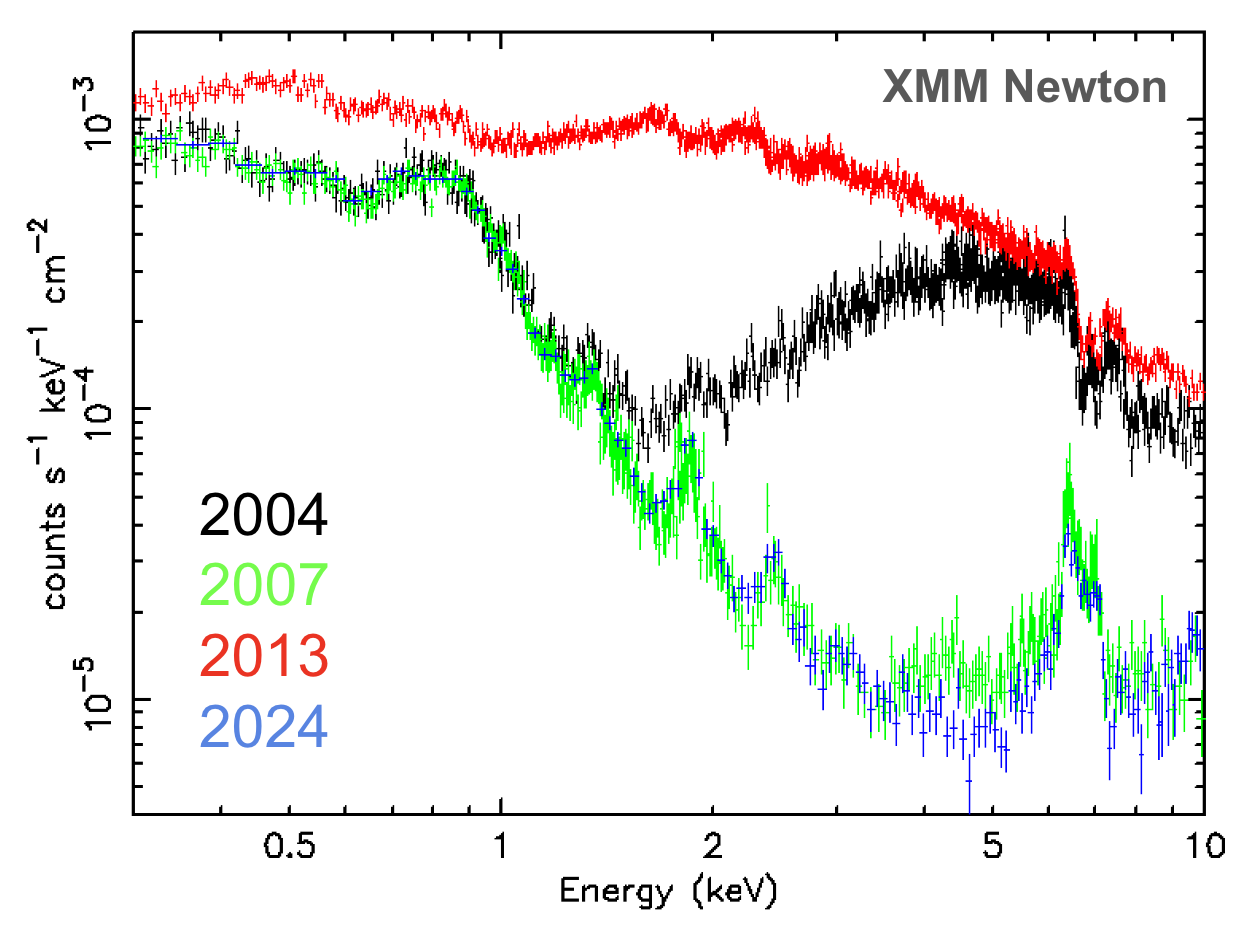}
    \caption{XMM-Newton EPIC-pn spectra of NGC 1365 from four epochs: 2004 (black), 2007 (green), 2013 (red), and 2024 (blue). The spectra illustrate long-term variability in the 0.3–10 keV band, with transitions between Compton-thin (e.g., 2013) and Compton-thick states (e.g., 2007, 2024). The source also undergoes transitions between these states on much shorter timescales of hours to days. The 2024 observation is contemporaneous with the first XRISM observation (Obs1) and captures the source in a heavily obscured state. All spectra are shown in the observed frame and have been binned for clarity.}
    \label{fig:XMM}
\end{figure}

\begin{figure}[t!]
    \centering
    \includegraphics[width=\columnwidth]{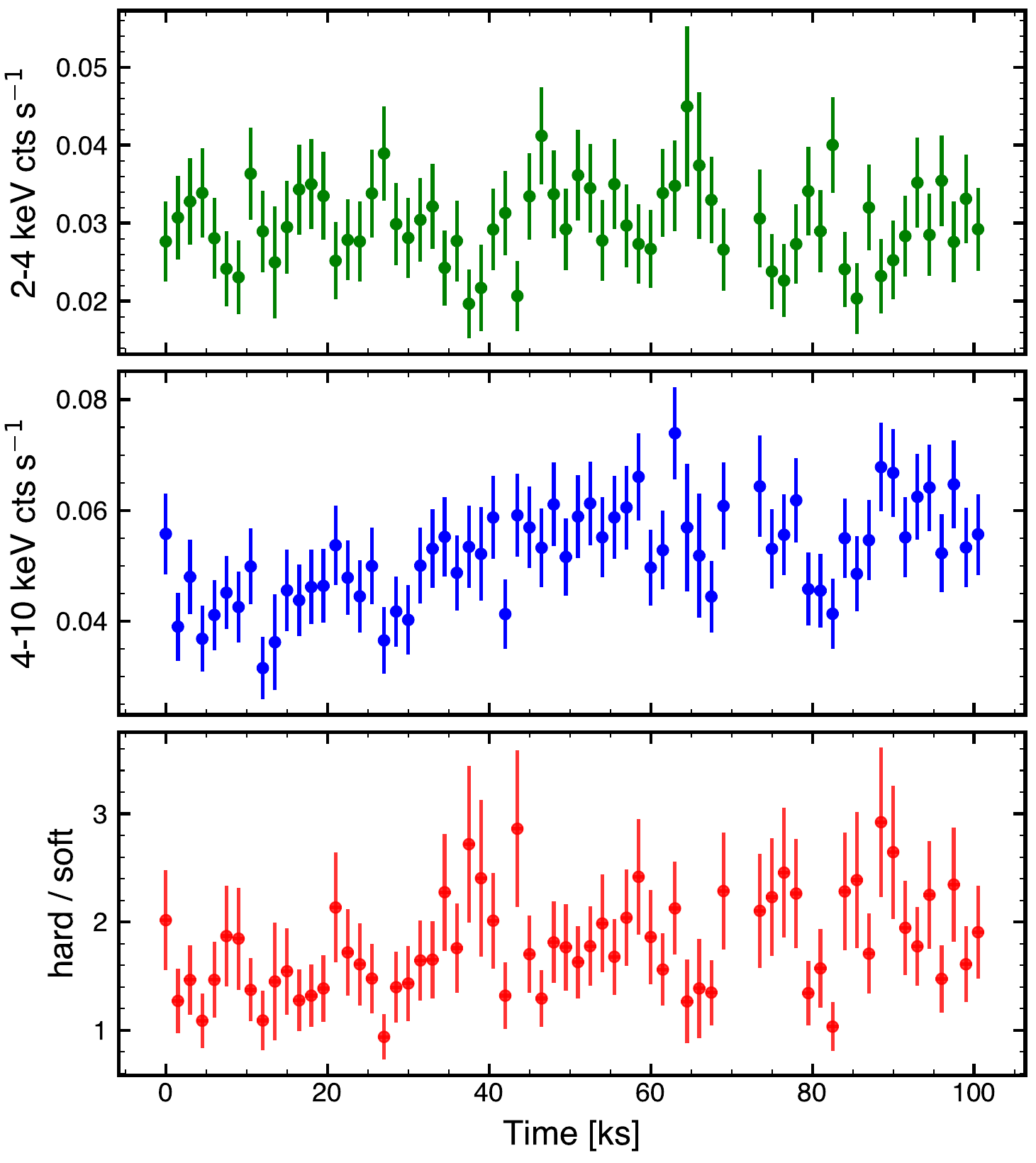}
    \caption{XMM-Newton EPIC-pn light curves of NGC 1365, simultaneous to XRISM/Resolve Obs1, in two energy bands: 2--4 keV (top, green) and 4--10 keV (middle, blue), and the corresponding hardness ratio (bottom, red) defined as the ratio of the 4--10 keV to 2--4 keV count rates. The light curves are binned at 1500 s. The hard band shows up to $\sim50\%$ variability over a $\sim$100 ks observation, suggesting the presence of a transmitted intrinsic continuum component even during this Compton-thick state.}
    \label{fig:Fig_lc_XMM.pdf}
\end{figure}

\begin{figure*}
\centering
\includegraphics[width=\textwidth]{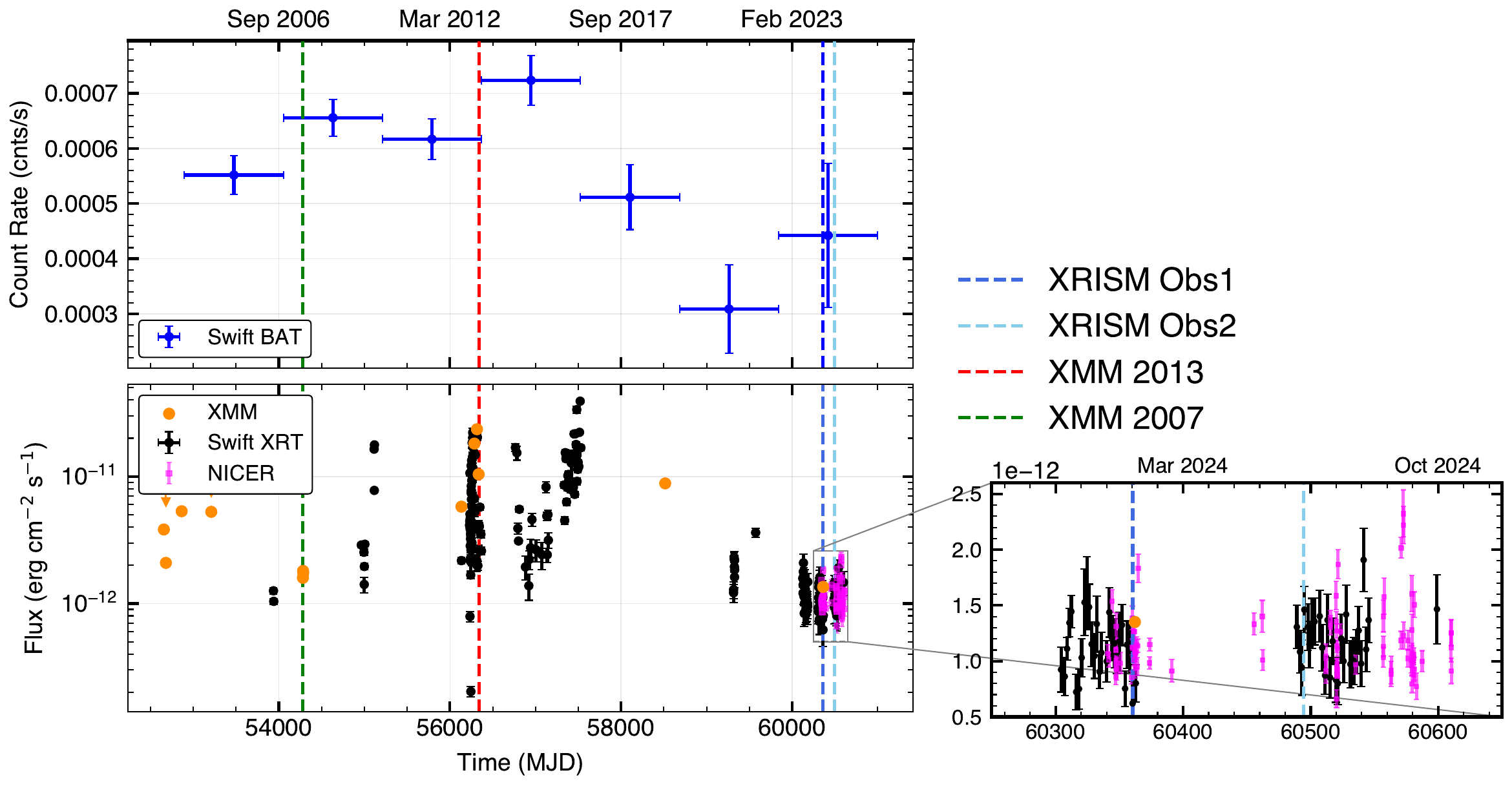}
\caption{Long-term X-ray light curves of NGC 1365. The top panel shows the Swift BAT 15--50~keV count rates averaged over $10^8$~s bins. The bottom panel displays the X-ray ($\sim$~0.3--10~keV) fluxes from XMM-Newton (orange), Swift XRT (black), and NICER (magenta). Vertical dashed lines mark the two XRISM observations (blue \& sky blue), as well as archival XMM-Newton observations from 2007 (green) and 2013 (red), corresponding to the spectra shown in Fig.~\ref{fig:XMM}. The right-hand panel shows a zoom-in of the NICER and Swift monitoring around the time of the XRISM observations. Together, these light curves capture the significant variability of the source and reveal a long-term decline in both soft and hard X-ray flux starting around 57500~MJD, leading to a persistent low-flux state.}
\label{fig:lc}
\end{figure*}

\noindent Fig.~\ref{fig:XMM} shows four XMM-Newton observations of NGC~1365 taken between 2004 and 2024. These illustrate the typical transitions between Compton-thick states, where the spectrum is dominated by distant reflection with a heavily absorbed power-law, and Compton-thin states, where the direct continuum emerges and a relativistically broadened iron line appears (e.g., \citealt{2013Natur.494..449R}; \citealt{2014ApJ...788...76W}; \citealt{2015MNRAS.446..737K}). While the hard X-rays vary dramatically, the flux below 1\,keV exhibits minimal variability, as it is dominated by emission from the galaxy’s circumnuclear star-forming region (e.g., \citealt{2009ApJ...694..718W}; \citealt{2013MNRAS.429.2662B}), except in the most unobscured state (red). In blue, we also include the XMM-Newton spectrum taken concurrently with our XRISM observation in Feb 2024 (Obs1). The 2024 observations of NGC 1365 were among the faintest seen from this source.

This absorption variability is a hallmark of NGC 1365, occurring on timescales ranging from days down to as short as hours. However, during our XRISM observations, NGC 1365 was found in an unprecedented, {\em persistent} low-flux state. Fig.~\ref{fig:lc} shows both the hard X-ray (Swift/BAT; 14-195~keV) and soft X-ray (Swift/XRT, NICER and XMM-Newton; 0.3--10~keV) light curves from 2007-2025. Up until $\sim 2018$, the source typically varies in soft X-rays by an order of magnitude and is consistently bright in hard X-rays ($\sim 0.00006$ counts~s$^{-1}$ or $4\times10^{-10}$~ergs~cm$^{-2}$~s$^{-1}$ in 14--195 keV), suggesting an intrinsically bright continuum. However, starting somewhere between 2016--2018, there began a sustained decline in soft X-ray flux, from  $ >10^{-11}~\mathrm{erg\,s^{-1}\,cm^{-2}}$ ($L \approx 10^{42}~\mathrm{erg\,s^{-1}}$) to $9.5 \times 10^{-13}~\mathrm{erg\,s^{-1}\,cm^{-2}}$ ($L \approx 6.7 \times 10^{40}~\mathrm{erg\,s^{-1}}$) in July 2024. The steady decrease was seen both in soft and hard X-rays indicating that NGC 1365 has become intrinsically fainter. This low-flux state continues through the time of the XRISM observations, during which we measure an observed flux of around $1.8 \times 10^{-12}~\mathrm{erg\,s^{-1}\,cm^{-2}}$ ($L_{\text{obs}} \approx 1.2 \times 10^{41}~\mathrm{erg\,s^{-1}}$) in the combined observation between 0.3 and 10~keV (using the simultaneous PN observation to extend down to 0.3~keV). Although the reduced X-ray flux results in diminished signal-to-noise, it offers a valuable opportunity to study this AGN in an intrinsically low-luminosity state, and under conditions of heavy obscuration, while simultaneously demonstrating XRISM's high-resolution capabilities even for faint sources.

\begin{figure*}[t]
\vspace*{-5mm} 
\includegraphics[width=\textwidth]{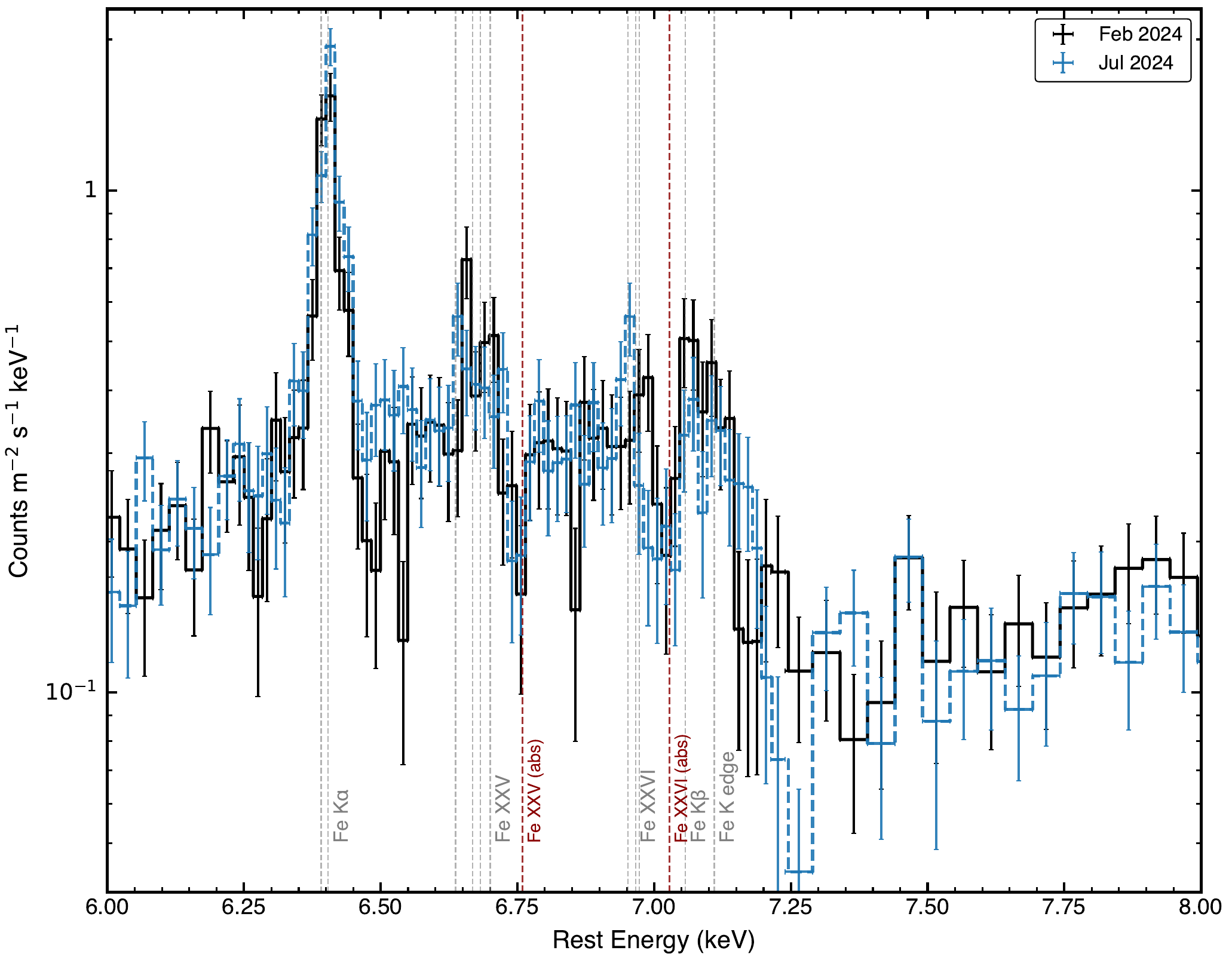}
\caption{XRISM/Resolve spectra of NGC 1365 in the 6.0--8.0~keV energy range from two epochs: February 2024 (black) and July 2024 (blue). The vertical dashed lines indicate the expected rest-frame energies of key emission and absorption features, including the Fe K$\alpha$ fluorescence line at 6.4~keV, Fe\,\textsc{xxv} lines at 6.70--6.75~keV, Fe\,\textsc{xxvi} Ly$\alpha$ lines around 6.95--7.05~keV, and the Fe K$\beta$ and K-edge features near 7.1~keV. Outside this energy range, the spectra are largely featureless aside from known non--X-ray background (NXB) lines. The two epochs show no striking visual differences, and the data have been binned for clarity.
}
\label{fig:resolve}
\end{figure*}

Fig.~\ref{fig:resolve} shows both of the XRISM/Resolve observations spectra from Feb and Jul 2024. Despite being separated by nearly five months, the spectra are strikingly similar in both flux and overall spectral shape, suggesting the source may have remained in this stable faint state throughout.  

Overall, the 2--10 keV Resolve spectra are featureless outside the 6–8 keV range, aside from narrow lines attributed to the non-X-ray background. The spectrum exhibits a prominent Fe K$\alpha$ emission line complex at $\sim$6.4 keV in the rest frame, along with Fe K$\beta$ at $\sim$~7.1 keV. The spectrum also reveals strong Fe\,\textsc{xxv} He$\alpha$ and Fe\,\textsc{xxvi} Ly$\alpha$ absorption and emission lines between 6.70 and 7.0 keV. A dip around 7.1 keV is also present, corresponding to a strong iron K-edge. There does not appear to be variability in the Fe\,\textsc{xxv} and Fe\,\textsc{xxvi} emission or absorption features between observations, and so for the following spectral analysis of the ionized winds (Section~\ref{subsec:pion}), we combine the two XRISM datasets to maximize signal. Combining observations yields a total exposure of $\sim$~450~ks.

\begin{figure}[]
    \centering
    \includegraphics[width=\columnwidth]{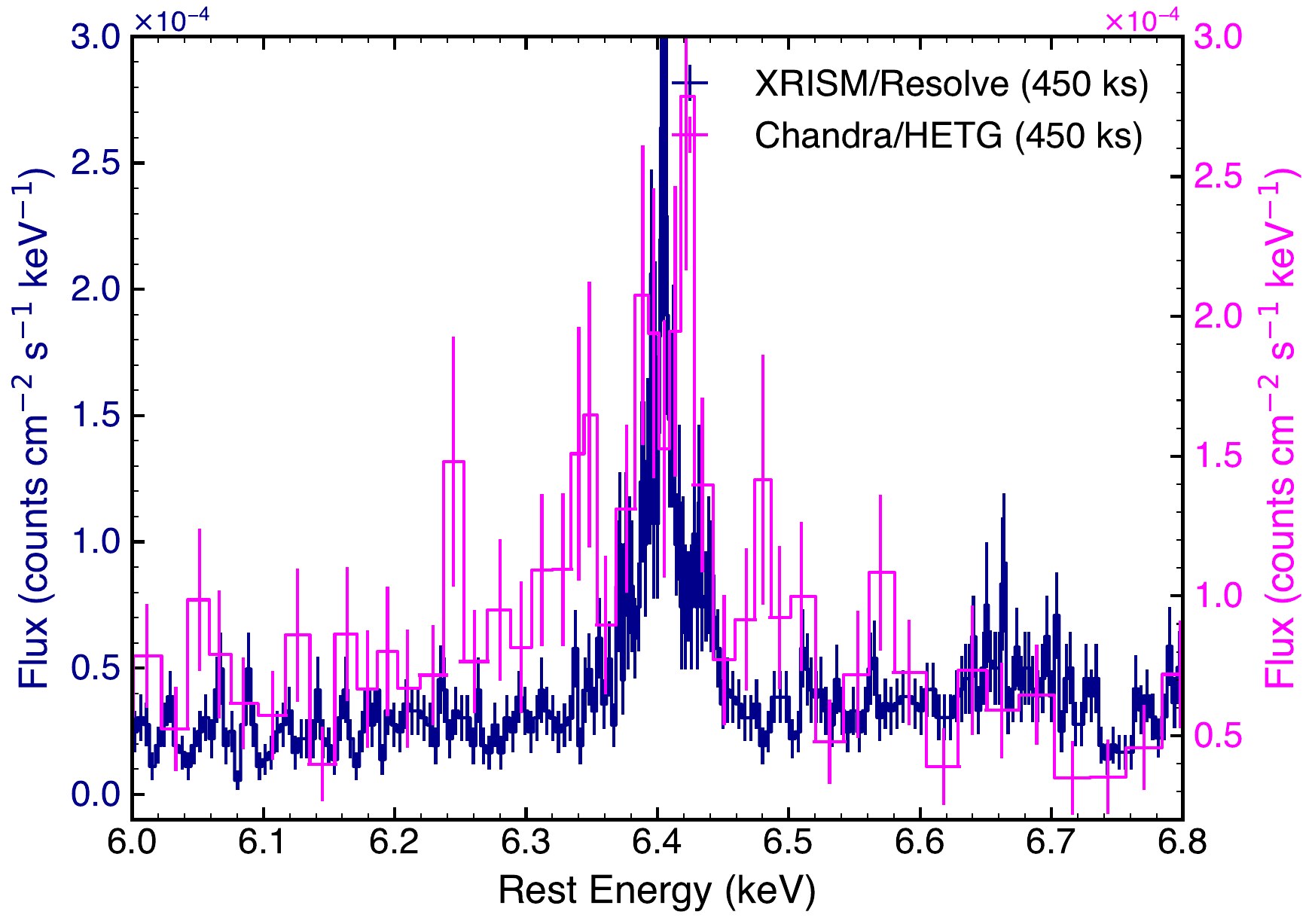}
    \caption{Comparison of the Fe K$\alpha$ emission region in NGC 1365 as observed by XRISM/Resolve (dark blue, left axis) and Chandra/HETG (magenta, right axis), each with an effective exposure of $\sim$~450~ks. The Fe K$\alpha$ fluorescence line at 6.4~keV is more clearly resolved in the XRISM data, revealing evidence for intrinsic broadening. This comparison demonstrates XRISM’s enhanced spectral sensitivity and resolution for NGC 1365, even during a relatively faint state of the source.}
    \label{fig:chandra}
\end{figure}

Zooming into the Fe K$\alpha$ region peaking at 6.4~keV reveals that the line is broadened, as seen also in  NGC 4151 \citep{2024ApJ...973L..25X}. In Fig.~\ref{fig:chandra}, we compare a combined Chandra/HETG spectrum totaling 450~ks with the combined XRISM Resolve spectrum of the same exposure, to illustrate XRISM’s enhanced capabilities despite the persistent faint/Compton-thick state. Earlier Chandra/HETG observations of NGC 1365 in a Compton-thick state \citep{2015MNRAS.453.2558N} found evidence at $>99$ per cent confidence that the Fe K$\alpha$ was resolved with a width of $\sim 3000$~km~s$^{-1}$, and now, with XRISM, we confirm this result clearly. The Fe K$\alpha$ line will be discussed more in Section~\ref{subsec:Fe}.

\section{Spectral Modeling Results} \label{sec:modeling}

\noindent We perform spectral modeling of the combined XRISM observations using the X-ray spectral fitting package \textsc{SPEX} v3.08.01 (\citealt{1996uxsa.conf..411K}). We use Cash statistics (\citealt{1979ApJ...228..939C}) to assess the goodness of fit, and all uncertainties are reported at the 1$\sigma$ confidence level. We assess the significance of additional spectral components by interpreting the change in C-statistic ($\Delta C$) for the corresponding change in degrees of freedom via the $\chi^2$ distribution. While this is standard in the field, we note that this is an asymptotic approximation that can yield optimistic sigma significances in the low-count regime (e.g, \citealt{2002ApJ...571..545P}). We assume a standard $\Lambda$CDM cosmology with $H_0 = 70$\,km s$^{-1}$\,Mpc$^{-1}$, $\Omega_{\rm m} = 0.3$, and $\Omega_{\Lambda} = 0.7$, corresponding to a luminosity distance of 23.5 Mpc for NGC~1365. Spectra are optimally binned, for modeling, following the prescription of \citet{Kaastra16}.

We model the 2--10 keV continuum with a cutoff power-law ({\tt pow $\times$ 2etau}) characterized by a best-fit photon index of $\sim 2$. The upper energy is fixed at 300~keV, typical of AGN X-ray coronae (\citealt{2015MNRAS.451.4375F}) while the lower energy cut-off is set to the Lyman limit. Freeing the energy limits does not significantly change the fit. 

The spectrum is obscured by a partial covering absorber, which has a column density of $\sim 1.4\times10^{24}$ cm$^{-2}$ and a covering factor of $\sim$98\%, required to reproduce the curvature of the spectrum in the 2--4 keV band. This partial covering absorption is modeled using the SPEX model \texttt{hot} \citep{2004A&A...423...49D, 2005A&A...434..569S}, which attenuates the continuum and accounts for the Fe K edge. When the electron temperature parameter is set to its default low value ($8 \times 10^{-6}$ keV), the model effectively mimics the transmission of a neutral plasma, similar to the \texttt{zpcfabs} model in XSPEC. In this case, the {\tt hot} model prefers to be either slightly ionized with an effective electron temperature between 0.001--0.004~keV corresponding to $\sim$ Fe\,\textsc{iv} to Fe\,\textsc{vi} (40,000-50,000~K) or outflowing  at  $-1500$~km~s$^{-1}$. Allowing both temperature and velocity to vary introduces significant degeneracy, making it difficult to disentangle the two effects. To mitigate this, the fits presented in this work fix the partial covering absorber velocity to zero and allow only the temperature of the SPEX {\tt hot} model to vary. Additionally, the high covering fraction associated with the partial covering absorber may instead reflect electron scattering that is not accounted for in the current model. We also assume that this partial covering absorber lies beyond the photo-ionized absorber and emitter.  

Because a relativistic line has been seen in this source before \citep{2013Natur.494..449R}, we attempted a fit assuming an intrinsically faint X-ray source, and a relativistically broadened iron line to account for the curvature at 4--5~keV, but the fit statistic was significantly worse. It is worth noting that previous studies using XMM-Newton, Suzaku, and NuSTAR (e.g, \citealt{2013MNRAS.429.2662B},\citealt{2014ApJ...788...76W}) have established the presence of relativistic reflection in NGC~1365, even under variations in the absorbing column by factors of $\sim$100. A broadband model incorporating a physical reflection component fit to the XMM-Newton, XRISM and NuSTAR data will be explored in more detail in future work. 

We also include Galactic absorption fixed at a column density of 0.012$\times10^{22}$ cm$^{-2}$ \citep{2016A&A...594A.116H}, using a second {\tt hot} model. We account for the non-X-ray background using a SPEX file model based on the extracted NXB spectrum (see section~\ref{subsec:xrism}).

Although the intrinsic continuum cannot be fully constrained without broadband modeling, we estimate an intrinsic (i.e., absorption-corrected) 2--10~keV luminosity of $\sim 3.1\times10^{42}~\mathrm{erg~s}^{-1}$ using the XRISM/Resolve data alone.

\subsection{Modeling of the Absorption and Emission Lines} \label{subsec:pion}

\begin{figure}[]
    \centering
    \includegraphics[width=\columnwidth]{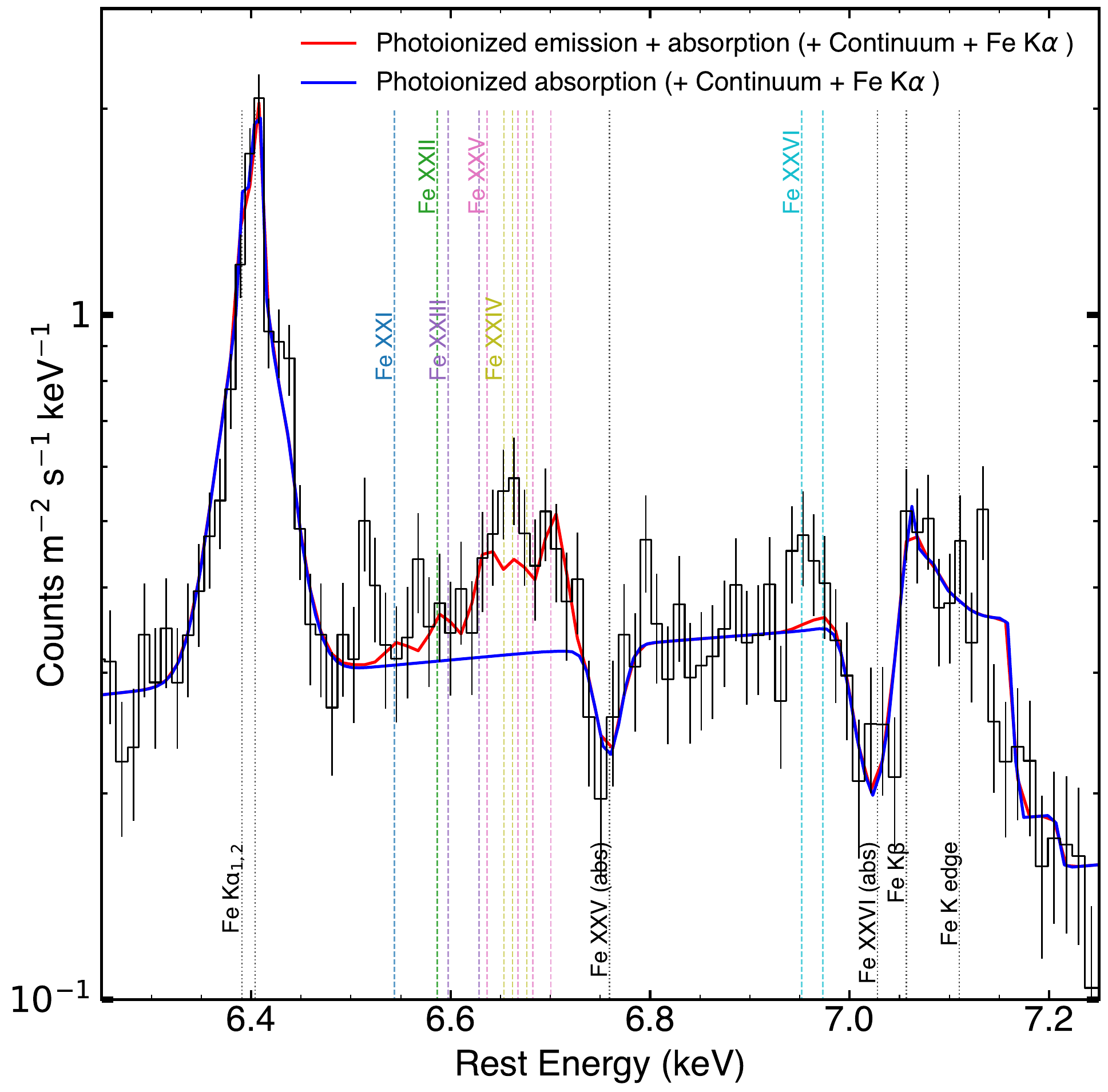}
    \caption{Best-fit model for the XRISM/Resolve combined spectrum in the region of interest, assuming both the emission and absorption arise from photoionized gas. The full model is shown in red, with the photoionized absorption component plotted with the blue model. The strongest lines contributing to the photoionized emission component of the model are indicated with colored labels and dashed lines. The emission consists of neutral and ionized iron lines, with a weak contribution from Fe\,\textsc{xxvi} near 7.0~keV.}
    \label{fig:pion}
\end{figure}

\noindent We resolve the Fe\,\textsc{xxv} and Fe\,\textsc{xxvi} absorption lines, and find them blueshifted by approximately $-2500$ km s$^{-1}$ with a root mean square velocity of $\sim600$ km s$^{-1}$. We model the absorption lines with a {\tt pion}  photoionized absorption component \citep{2016A&A...596A..65M} that has a column density of $10^{23}$ cm$^{-2}$ and an ionization parameter of $\log(\xi$ [~erg~cm~s$^{-1}$]) $\sim3.8$. The covering factor is fixed to unity in our fits to mitigate degeneracy with the column density. Similar highly ionized absorption lines (at velocities roughly 2000-5000~km~s$^{-1}$) were seen previously in CCD observations with XMM-Newton/PN and in Suzaku (e.g., \citealt{2005ApJ...630L.129R, 2013MNRAS.429.2662B}). 

For the first time in this source, we detect emission lines in the 6.6--6.7 keV range, associated with Fe\,\textsc{xxv} transition and other lower ionization Fe lines. First, we attempt to model these emission lines as P Cygni-like features from the same photoionized gas producing the absorption lines described above. We use a {\tt pion} photoionized emission component, and find an ionization parameter of log($\xi$~[erg cm s$^{-1}$])$\sim$3 and column density of $\sim 2\times10^{21}$ cm$^{-2}$. Including this emission component improves the fit significantly by $\Delta C = 48$ for 4 d.o.f. Both the Fe K$\alpha$ emission line and the {\tt pion} emission component require a blueshift of approximately 150-190 km s$^{-1}$, either adjusting for an offset in the assumed cosmological redshift or a shared intrinsic velocity suggesting a common origin of the Fe K line and the ionized emission. The photo-ionized emission model component also fits a S\,\textsc{xvi} line at around 2.62~keV. There is a hint of Fe\,\textsc{xxvi} as well in the data, we find that adding a Gaussian at its average transition energy improves the fit by a marginal significance of 2.2$\sigma$.

The strongest feature in the spectrum is the Fe K$\alpha$ emission at $\sim$6.4~keV. The line complex cannot be fitted with a single component. We adopt the experimental intrinsic line profiles from \citet{1997PhRvA..56.4554H}: H97 hereafter, for the K$\alpha$ and K$\beta$ transitions. We fit a narrow core component, and a second Hölzer profile that is broadened with a Gaussian smoothing kernel ({\tt vgau}), we measure a velocity width $\sigma$ of around 1300 km s$^{-1}$. When the narrow core is allowed to broaden through the inclusion of an additional Gaussian smoothing component, we measure a velocity broadening of $v \sim 100 \pm 50\,\mathrm{km\,s^{-1}}$. This modest broadening of the narrow core does not affect the measured width of the broad component. Based on visual inspection, we allow both the narrow and broadened Hölzer components to share a common velocity shift using the {\tt reds} model, accounting for a deviation from the cosmological redshift or potential bulk motion of the emitting gas. In the next section, we discuss more physical models for the broadening, but for the purposes of fitting the 2--10~keV spectrum to characterize the outflows, here we simply use a phenomenological gaussian to fit the broad Fe~K$\alpha$ line.

The full model fitting the 2--10~keV {\em Resolve} spectrum in SPEX syntax is: 

\begin{equation} \label{eq:model1}
  \begin{array}{l}
    {\tt \{ pow \times 2etau \times pion\_emis \times pion\_abs \times hot } \\
    {\tt + (H97 + H97 \times vgau) \times reds \} \times reds \times hot + nxb }
  \end{array}
\end{equation}

\begin{table}[t!]
\centering
\caption{Best-fitting photoionization model parameters for the combined XRISM Resolve spectrum\label{tab:pion}}
\footnotesize
\begin{tabular}{llc}
\hline
Component & Parameter & Value \\
\hline
\textsc{pow} & $\Gamma$ & $1.95^{+0.11}_{-0.08}$ \\
             & Norm ($10^{49}$ ph s$^{-1}$ keV$^{-1}$) & $115^{+39}_{-23}$ \\
\hline
\textsc{hot} & $N_\mathrm{H}$ ($10^{22}$ cm$^{-2}$) & $137^{+6}_{-5}$ \\
             & $t$: elec. temp. (keV) & $(3.83^{+0.35}_{-0.26}) \times 10^{-3}$ \\
             & $f_{\rm cov}$ & $0.984^{+0.003}_{-0.002}$ \\
\hline
\textsc{pion (abs.)} & $N_\mathrm{H}$ ($10^{22}$ cm$^{-2}$) & $10.8^{+7.3}_{-3.6}$ \\
                     & $\log \xi$ (erg cm s$^{-1}$) & $3.78^{+0.25}_{-0.19}$ \\
                     & $f_{\rm cov}$ & 1\tablenotemark{a} \\
                     & $\sigma_v$ (km s$^{-1}$) & $580^{+160}_{-120}$ \\
                     & $v_{\text{out}}$ (km s$^{-1}$) & $-2600 \pm 160$ \\
\hline
\textsc{pion (em.)} & $N_\mathrm{H}$ ($10^{22}$ cm$^{-2}$) & $0.21^{+0.05}_{-0.06}$ \\
                    & $\log \xi$ (erg cm s$^{-1}$) & $2.97^{+0.15}_{-0.11}$ \\
                    & Covering Factor $\Omega$ & 1\tablenotemark{a} \\
                    & $\sigma_v$ (km s$^{-1}$) & $447^{+133}_{-114}$ \\
                    & $v_{\text{out}}$ (km s$^{-1}$) & $-185^{+116}_{-136}$ \\
\hline
\textsc{H97 (x2)} & Norm (narrow) (arb. units) & $14.4 \pm 2.5$ \\
            & Norm (broad) (arb. units) & $41.2 \pm 3.5$ \\
            & $\sigma$ (broad) (km s$^{-1}$) & $1255^{+122}_{-111}$ \\
            & $z$ (shared) ($10^{-4}$) & $-4.82^{+0.92}_{-0.97}$ \\
\hline
& C-stat/d.o.f. & $2112/2120$ \\
\hline
\end{tabular}
\tablenotetext{a}{Fixed parameter}
\end{table}

Fig.~\ref{fig:pion} shows the best-fitting model assuming photoionized emission, and Table~\ref{tab:pion} summarizes the values and uncertainties of the main parameters.

An alternative explanation for the observed emission lines is that they originate from a collisionally ionized plasma rather than a photoionized medium. Collisionally ionized emission can arise from shocks, such as AGN-driven outflows interacting with surrounding gas, or from hot plasma associated with nuclear star formation. NGC~1365 is a known starburst galaxy (e.g. \citealt{1980A&A....87..245V}; \citealt{1996A&A...305..727H}) and soft X-ray emission lines attributed to collisionally ionized plasma have previously been detected in this source, particularly below 2 keV in Chandra and XMM-Newton observations (\citealt{2009ApJ...694..718W}, \citealt{2015MNRAS.453.2558N}). To explore whether the observed emission lines are consistent with a model of thermal plasma, we use the same setup as in Equation~\ref{eq:model1}, but replace the emission component with a collisional ionisation equilibrium model ({\tt cie} in SPEX). The full model in SPEX syntax is:

\begin{equation} \label{eq:model2}
  \begin{array}{l}
    {\tt \{ pow \times 2etau \times pion\_abs \times hot + cie \times reds } \\
    {\tt + (H97 + H97 \times vgau) \times reds \} \times reds \times hot + nxb }
  \end{array}
\end{equation}

\begin{figure}[]
    \centering
    \includegraphics[width=\columnwidth]{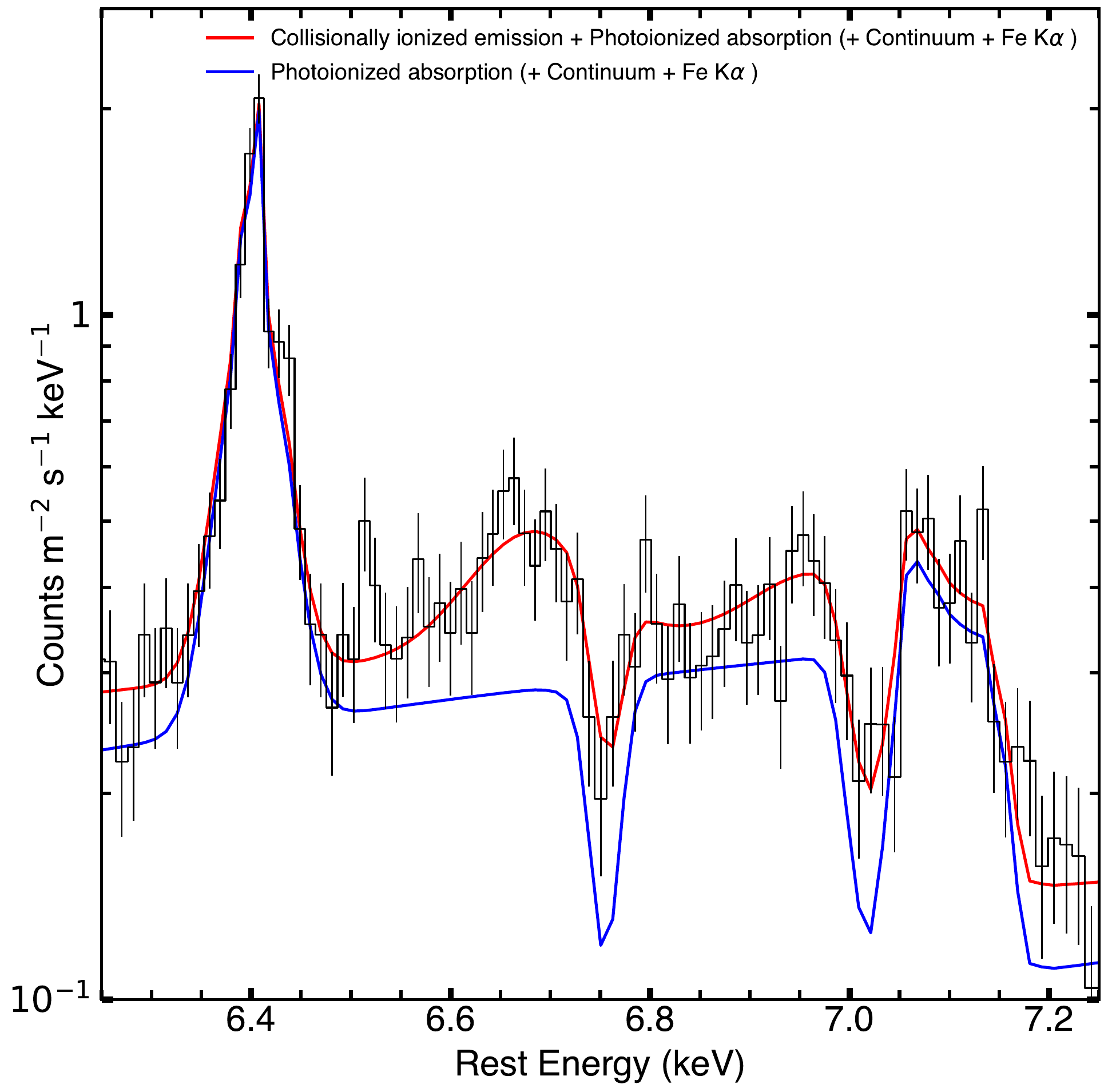}
    \caption{Best-fitting model to the XRISM/Resolve spectrum assuming the observed ionized emission lines originate in a collisionally ionized plasma. The total model, including a thermal emission component, is shown in red, while the model without ionized emission is shown in blue. This model provides a comparably good description of the data as the photoionized emission scenario.}
    \label{fig:cie}
\end{figure}

\begin{table}[t!]
\centering
\caption{Best-fitting collisional ionization model parameters for the combined XRISM Resolve spectrum\label{tab:cie}}
\footnotesize
\begin{tabular}{llc}
\hline
Component & Parameter & Value \\
\hline
\textsc{pow} & $\Gamma$ & $2.06^{+0.22}_{-0.10}$ \\
             & Norm ($10^{49}$ ph s$^{-1}$ keV$^{-1}$) & $154^{+171}_{-51}$ \\
\hline
\textsc{hot} & $N_\mathrm{H}$ ($10^{22}$ cm$^{-2}$) & $142^{+12}_{-7}$ \\
             & $t$ (keV) & $(3.93^{+0.52}_{-0.28}) \times 10^{-3}$ \\
             & $f_{\rm cov}$ & $0.996^{+0.003}_{-0.004}$ \\
\hline
\textsc{pion (abs.)} & $N_\mathrm{H}$ ($10^{22}$ cm$^{-2}$) & $16^{+16}_{-6}$ \\
                     & $\log \xi$ (erg cm s$^{-1}$) & $3.91^{+0.55}_{-0.35}$ \\
                     & Covering Factor $\Omega$ & 1\tablenotemark{a} \\
                     & $\sigma_v$ (km s$^{-1}$) & $640^{+150}_{-120}$ \\
                     & $v_{\text{out}}$ (km s$^{-1}$) & $-2440^{+170}_{-140}$ \\
\hline
\textsc{cie (em.)} & Norm ($=n{_H}n{_e}V$) ($10^{69}$ m$^{-3}$) & $2.8^{+0.9}_{-0.8}$ \\
            & $t$ (keV) & $7.15^{+1.64}_{-1.15}$ \\
            & $v_{\rm rms}$ (km s$^{-1}$) & $2450^{+740}_{-1170}$ \\
            & $z$ ($10^{-4}$) & $-0.4 \pm 20$ \\
\hline
\textsc{H97 (x2)} & Norm (narrow) (arb. units) & $14.5 \pm 2.4$ \\
            & Norm (broad) (arb. units) & $41.2 \pm 3.4$ \\
            & $\sigma$ (broad) (km s$^{-1}$) & $1250^{+120}_{-110}$ \\
            & $z$ (shared) ($10^{-4}$) & $-4.79^{+0.90}_{-0.98}$ \\
\hline
& C-stat/d.o.f. & $2113/2120$ \\
\hline
\end{tabular}
\tablenotetext{a}{Fixed parameter}
\end{table}

Fig.~\ref{fig:cie} shows the resulting best fit which has a plasma temperature of $\sim$7.2~keV. Table~\ref{tab:cie} shows a summary of the key fit parameters. The fit statistic is comparable to that of the photoionized model. The parameters of the continuum and photoionized absorption models, which are jointly fitted alongside the {\tt cie} emission component, remain consistent within $1\sigma$ uncertainties relative to the photoionized emission case, with the exception of the covering fraction associated with the partial covering absorber, which is slightly higher in the collisional ionization fits, exceeding a $2\sigma$ difference.

Statistically, we cannot determine if the emission lines are due to collisionally or photoionized emission. In Section~\ref{sec:discussion}, we discuss tests for distinguishing between these scenarios to ultimately determine their origin.

\subsection{Disk Broadening of the Fe K$\alpha$ Line} \label{subsec:Fe}

\begin{figure*}[t]
    \centering
    \includegraphics[width=\textwidth]{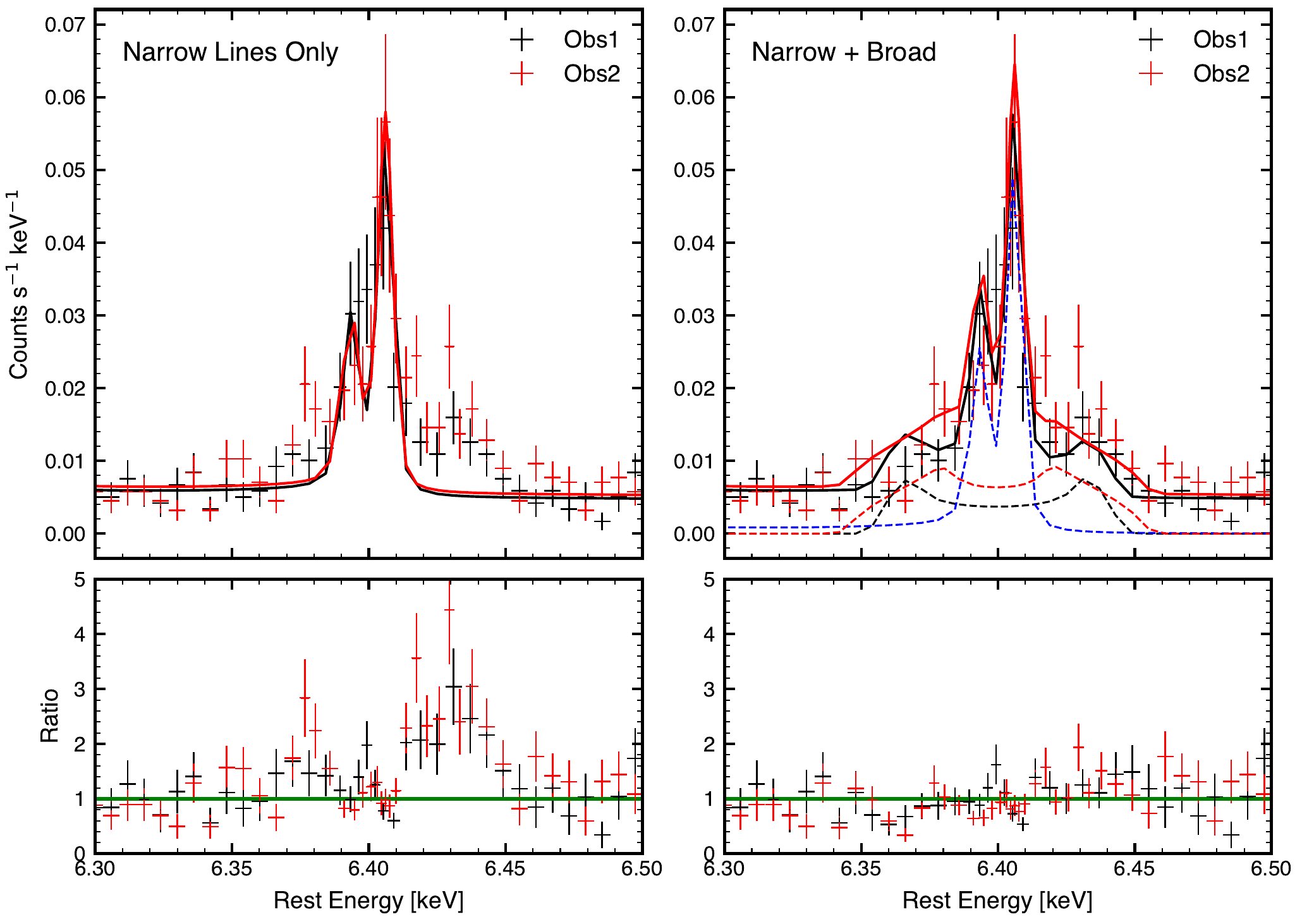}
    \caption{Fe K$\alpha$ line profiles from the two XRISM/Resolve observations of NGC 1365 (Obs1: black; Obs2: red). The left panel shows fits using only narrow line components modeled with the Hölzer profile. The residuals (data/folded model) reveal clear broadening and asymmetry in the wings of the line (the residuals in the blue wing are larger than the red wing). The right panel shows fits including both narrow and disk-broadened components (\texttt{diskline}); dashed lines represent the individual model components. Including a broad component significantly improves the fit, although some asymmetry and residual structure remain unaccounted for in the ratio panel.}
    \label{fig:Fe}
\end{figure*}

\noindent Earlier Chandra/HETG observations of NGC~1365 showed that the Fe~K$\alpha$ complex was best fit by a $\text{FWHM of } \sim3000^{+1900}_{-1350}$ ($\sigma = 1300^{+800}_{-600})~\mathrm{km\,s}^{-1}$ \citep{2015MNRAS.453.2558N}. Now XRISM confirms this broad component, in addition to a narrow core, and provides tighter constraints on its width. To explore the origin of the broad component, we fit the $6.1$--$6.6$~keV band, chosen to encompass the full structure of the Fe K$\alpha$ line. While the 2--10~keV spectrum modeled in Section~\ref{subsec:pion} was optimally binned \citep{Kaastra16}, we chose to bin this region only up to three times the instrument resolution to preserve as much structure as possible. Given the apparent visual differences in this region between the two observations, we opt to model Obs1 and Obs2 individually rather than using a combined spectrum. 

We use the relativistic disk line broadening model {\tt diskline} \citep{1989MNRAS.238..729F} added to a Hölzer line profile (Fe K$\alpha$ complex) for the narrow core, all superimposed on an underlying power-law continuum. 
 The main parameters of {\tt diskline} are the inner radius ($r_{\mathrm{in}}$), outer radius ($r_{\mathrm{out}}$), disk inclination angle ($i$), and the radial emissivity index ($q$), where the disk emissivity is assumed to follow $I(r) \propto r^{q}$. Using trial and error, we determine that we are not able to accurately constrain the inclination given the quality of the data and the degeneracies in the parameter space. Thus, the inclination is tied between the two observations and is fixed to 60$\degree$. This is the inclination estimated for inner accretion disk as inferred from the relativistic line seen previously in an unobscured state, so we are assuming that the inner disk and outer disk/BLR are aligned (e.g, \citealt{2014ApJ...788...76W}). 

We find the emissivity index to be loosely constrained at $q = -2.6^{+1.2}_{-0.9}$. The inner and outer radii are $r_{\mathrm{in,\,Obs1}} = (1.5^{+0.3}_{-0.6}) \times 10^4$ and $r_{\mathrm{out,\,Obs1}} = (3.3^{+0.3}_{-0.6}) \times 10^4$ for Obs1, and $r_{\mathrm{in,\,Obs2}} = (1.0^{+0.3}_{-0.6}) \times 10^4$ and $r_{\mathrm{out,\,Obs2}} = (9^{+19}_{-4}) \times 10^4$ for Obs2, all in units of $GM/c^2$. In this setup, the inner radii are statistically consistent within 1$\sigma$ across the two observations, while the outer radii differ at the $\sim$1.4$\sigma$ level, indicating marginal consistency. The C-statistic versus outer radius curve in Obs2 flattens beyond the 1$\sigma$ upper bound, and the 2$\sigma$ interval fails to converge, suggesting the quoted value effectively serves as a lower limit. These inferred distances indicate that the emission region of the broad Fe K$\alpha$ line extends across the typical $10^4\ R_g$ scale of the AGN BLR. 

Fig.~\ref{fig:Fe} shows the best-fitting model (C-stat/d.o.f$=766/663$) and line profiles, with and without the broad component. The residuals to the narrow-only fit show that the broad line is asymmetric: the blue wing of the line is more prominent than the red wing. In particular, the broad component in Obs1 appears to be double peaked, and indeed is best fit by a double-peaked {\tt diskline} model, which leads to a tighter constraint on the outer radius of the accretion disk (relative to Obs2). While the {\tt diskline} model in Fig.~\ref{fig:Fe}-{\em right} is statistically a good fit to the data, the residuals do suggest some additional structure not captured by the model. For both observations, when fitting with this largely symmetric model, the red wing of the line is over-predicted while the blue wing is under-predicted, indicating some asymmetry in the line. This could suggest a different geometry of the emitting region, or that our assumption of simply neutral gas is incorrect (e.g., \citealt{2004ApJS..155..675K}). Emission from mildly ionized material will be explored in more detail in future work. 

\section{Discussion} \label{sec:discussion}

\subsection{The Nature of the Obscuration in NGC 1365} \label{sec:unification}

\noindent Standard AGN unification models attribute the obscuration in Seyfert 2 galaxies to a parsec-scale torus (e.g., Circinus; XRISM Collaboration, in prep). However, in NGC 1365, the historical rapid variability in the line-of-sight column density on timescales of hours to weeks (e.g., \citealt{2005ApJ...623L..93R}, \citealt{2013MNRAS.429.2662B}) suggests that the obscuring material resides much closer to the central black hole. This has been interpreted as due to comet-like clouds in the broad-line region (e.g., \citealt{2010A&A...517A..47M}). The simultaneous presence of outflow signatures and emission lines, which are more typical of Seyfert 1 galaxies, alongside Compton-thick obscuration associated with Seyfert 2 sources, highlights the composite nature of NGC 1365.

In contrast to the short-term X-ray obscuration events previously reported, our XRISM observations occur within a sustained obscured state in the X-rays (Fig.~\ref{fig:lc}). The two XRISM/Resolve spectra, taken nearly five months apart (Feb and Jul 2024), are similar in flux and shape (Fig.~\ref{fig:resolve}), indicating that the source may have remained in a stable Compton-thick state. 

Two scenarios may explain the observed state: a new phase of obscuration or an intrinsically fainter continuum, perhaps due to a lower accretion rate. If due to obscuration, this would indicate a phenomenon distinct from the small comet-like BLR clouds previously observed in \citet{2010A&A...517A..47M}, such as a large cloud suddenly covering our line of sight, potentially at larger distances than the BLR. Alternatively, an intrinsic continuum decrease can explain the decline in overall flux (even at hard X-rays $>15$~keV). This would mean less radiation pressure on dusty gas, allowing obscuring material to accumulate and persist along the line of sight, consistent with the picture where less luminous AGN tend to be more heavily obscured (\citealt{2014MNRAS.441.3622R}; \citealt{2014MNRAS.440.3503C}; \citealt{2021RAA....21..199L}). The decrease in hard X-ray luminosity (Fig.~\ref{fig:lc}), together with the observed short-timescale variability in the 4–10 keV simultaneous XMM-Newton light curve, suggests that the current low-flux state is characterized by substantial line-of-sight obscuration, with a contribution from the intrinsic transmitted continuum.

Additional evidence that the obscuring material is not due to a classical neutral and uniform torus comes from its ionized (t $\sim$0.004~keV from the {\tt hot} model) and possibly outflowing nature (v $\sim-1500$~km~s$^{-1}$), as indicated by the blueshifted Fe~K edge at $\sim$7.1 keV seen in Fig.~\ref{fig:resolve}, and detailed in Tables~\ref{tab:pion} and \ref{tab:cie}. 

\subsection{Origin of the Ionized Absorption Lines} \label{sec:origin-lines}

\noindent This is the first time absorption lines are observed in a Compton-thick AGN among XRISM PV targets. The outflow is significantly detected despite the extreme obscuration and is among the most persistent obscurer-type outflows observed in AGN.

We can estimate an upper limit on the distance to the outflow as $R < L_{\rm ion} / (N_{\rm H} \xi)$ by assuming that the thickness ($\Delta r$) of the absorbing layer is not larger than its distance from the X-ray source (i.e $\Delta r/r < 1$) and using $N_{\mathrm{H}}\sim n C_v \Delta r \sim C_v (L/\xi) (\Delta r / r^2)$ , where $C_v < 1$ is the volume filling factor. We use a weighted average of the best-fitting values from Table 2 and 3 treating them as two independent measurements of the outflow properties. Thus, we adopt a column density of $N_{\mathrm{H}} = (1.2 \pm 4.9) \times 10^{23}~\mathrm{cm}^{-2}$, an ionization parameter of $\xi = (6300 \pm 2900)$ erg cm s$^{-1}$, and a $1$--$1000$ Ryd ionizing luminosity of $L_{\text{ion}} = (1.1 \pm 0.3) \times 10^{43}$ erg~s$^{-1}$, inferred from our best-fitting model. This yields
\begin{align*}
R &< (1.5 \pm 0.9) \times 10^{16}~\mathrm{cm} \\
  &\simeq (2.2 \pm 1.5) \times 10^4~R_g \simeq (0.005 \pm 0.003)~\mathrm{pc} \\
  & < 3.7 \times 10^4~R_g \quad (0.008~\mathrm{pc} , 2.4 \times 10^{16}~\mathrm{cm})
\end{align*}
where $R_g = \frac{GM_{\rm BH}}{c^2}$ and $\log(M_{\mathrm{BH}}/M_\odot) = 6.65 \pm 0.09$ \citep[e.g.,][]{2017MNRAS.468L..97O,2022ApJS..261....2K}. The final inequality represents a single-value upper limit, obtained by adding the mean and its $1\sigma$ uncertainty.

We also estimate a minimum launch radius by equating the outflow velocity to the escape velocity at that distance ($R \geq \frac{2GM}{v_{\rm out}^{2}}$). This approximation introduces considerable uncertainty, since we observe only the projected component of the velocity vector. Adopting an outflow velocity of $v_{\rm out} = (2500 \pm 100)$ km~s$^{-1}$, we obtain
\begin{align*}
R &> (1.9 \pm 0.4) \times 10^{16}~\mathrm{cm} \\
  &\simeq (2.8 \pm 0.9) \times 10^4~R_g \simeq (0.006 \pm 0.001)~\mathrm{pc} \\
  & > 1.9 \times 10^4~R_g \quad (0.005~\mathrm{pc} , 1.5 \times 10^{16}~\mathrm{cm})
\end{align*}

Under these numerous and simplifying assumptions, the outflow is constrained to lie at a distance of approximately $1.9$--$3.7 \times 10^4~R_g$ or $1.5$--$2.4 \times 10^{16}~\mathrm{cm}$ from the black hole. This is comparable to the location of the BLR, usually a few $\times 10^4~R_g$ (e.g, \citealt{2011ApJ...738..147S}). This distance is roughly a factor of 10--100 larger than the ionization-based estimates of the distance to the outflow reported by \citet{2013MNRAS.429.2662B} (185--5400~$R_g$) and \citet{2005ApJ...630L.129R} (100--200~$R_g$), but is consistent with the variability-based constraints from \citet{2013MNRAS.429.2662B}, which place the outflow between 15,000 and 61,000~$R_g$ for NGC 1365. It is also in agreement with the distance estimate from \citet{2014ApJ...795...87B} ($10^4~R_g < R < 10^{16}~\mathrm{cm}$), derived using the same ionization and escape velocity arguments.

Most interestingly, the radius inferred for the distance of the outflow matches that inferred from the broad component of the Fe~K$\alpha$ line in Section~\ref{subsec:Fe}. It is possible that these two components originate from the same gas. Indeed, CLOUDY models for ionized obscurers along our line of sight produce blueshifted absorption lines and also predict some re-emission in the form of a broadened Fe~K$\alpha$ line \citep{2020ApJ...898..141D}. This interpretation is corroborated by the asymmetry of the broad component of the Fe~K line seen in Fig.~\ref{fig:Fe}-{\em left}. Additionally, in classical photoionized models, relatively cool BLR gas at \( T \sim 10^4\text{--}10^5~\mathrm{K} \) can coexist in pressure equilibrium with much hotter phases (\( T > 10^7~\mathrm{K} \)), as illustrated in Fig.~5.9 of \citet{2013peag.book.....N}. This multi-phase structure naturally explains the presence of highly ionized lines arising on BLR scales.

The kinetic luminosity of the outflow was calculated following the assumptions in \citet{2024ApJ...974...91Z}, which adopts the lower-limit prescription from \citet{2013IAUS..290...45K} and the upper-limit from \citet{2005A&A...431..111B}. Assuming a maximal volume filling factor $C_{\mathrm{v}} = 1$ and a reasonable covering factor $\Omega/4\pi = 0.25$, using the same $N_{\mathrm{H}}$, $\xi$, $L_{\mathrm{ion}}$, and $v_{\mathrm{out}}$ as above, together with the distance estimate $r \sim 2 \times 10^{16}~\mathrm{cm}$ and a bolometric luminosity $L_{\mathrm{bol}} \sim 10^{43.29}~\mathrm{erg\,s^{-1}}$ \citep{2022ApJS..261....2K}, we obtain $ L_{\mathrm{kin_{min}}} = 9.7 \times 10^{40}~\mathrm{erg\,s^{-1}}$ and $L_{\mathrm{kin_{max}}} = 7.2 \times 10^{40}~\mathrm{erg\,s^{-1}}$, corresponding to $L_{\mathrm{kin}}/L_{\mathrm{bol}} \sim 0.49\%-0.37\%$. The seemingly larger lower limit relative to the upper limit is due to the uncertainties in both estimates. In any case, the inferred kinetic luminosity is comparable to that derived from molecular gas (CO; \citealt{2021ApJ...913..139G}) and ionized gas (H$\alpha$; \citealt{2018A&A...619A..74V}), but remains below (or at the limit of) the fraction of bolometric luminosity typically required for impactful kinetic-driven feedback ($0.5$--$5\%$; \citealt{2010MNRAS.401....7H}). With a more realistic volume filling factor, e.g.\ $C_{\mathrm{v}} \sim 0.08$ as suggested by \citet{2005A&A...431..111B} for AGN, the contribution of NGC~1365’s wind to large-scale feedback is likely even more negligible. Long-term monitoring will be required to better constrain the duty cycle of NGC~1365’s outflow.

\subsection{Properties of the Fe K$\alpha$ Line}

\begin{figure}[]
    \centering
    \includegraphics[width=\columnwidth]{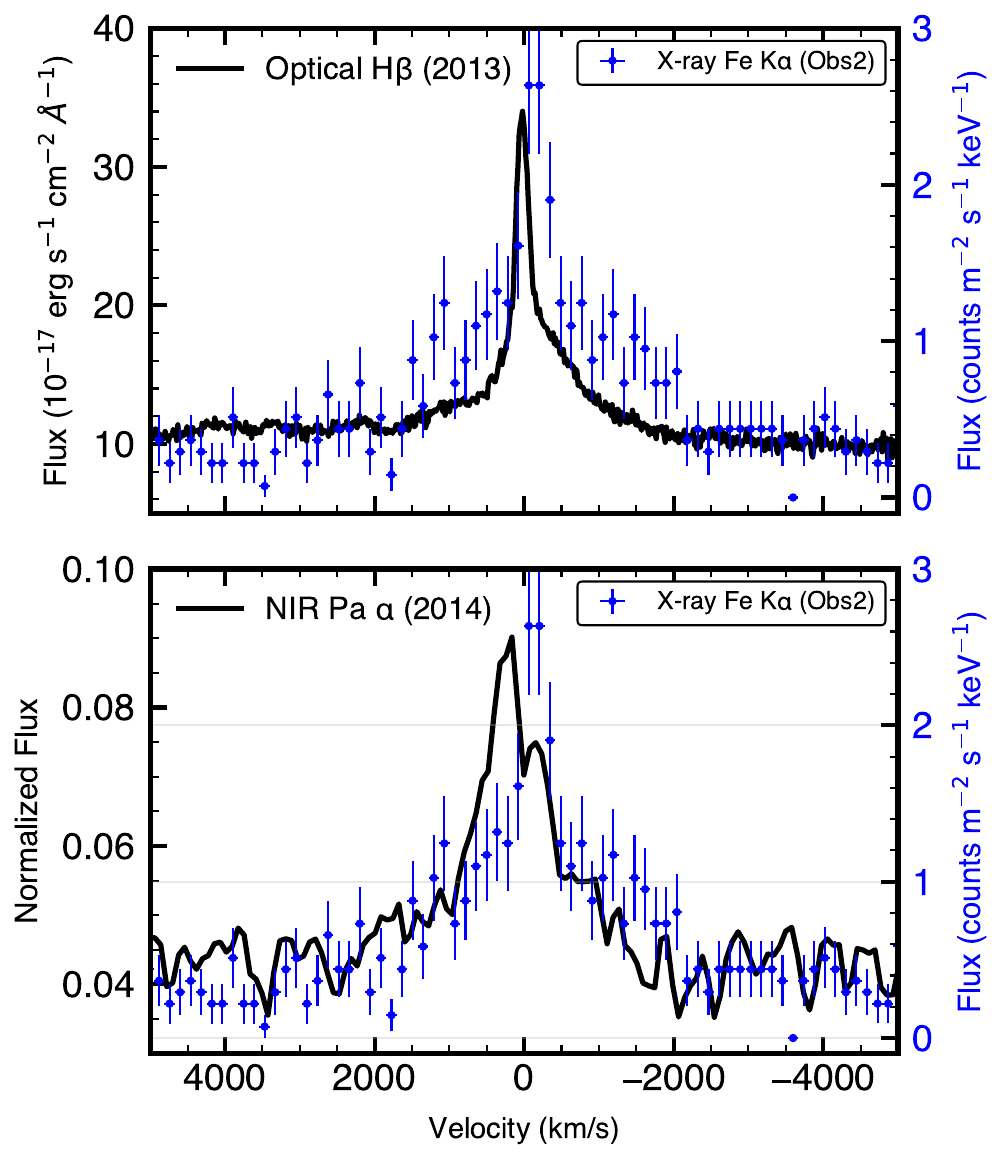}
    \caption{Overlay of the broad Fe K$\alpha$ line profile from XRISM/Resolve Obs2 (blue, right axis) with archival optical (H$\beta$, top) and near-infrared (Pa$\alpha$, bottom) broad emission lines (black). All spectra are plotted on a velocity axis centered on the rest frame of each respective line. The flux of the Pa$\alpha$ line has been normalized to its peak for visual comparison. The lines exhibit comparable broadening, consistent with an origin at similar physical scales.}
    \label{fig:opticalNIR}
\end{figure}

\noindent The origin of the Fe K$\alpha$ emission line in AGN has been widely debated. As noted by \citet{2004ApJ...604...63Y}, the line can plausibly arise in three separate sites: the accretion disk, the torus, and the BLR. In most AGN, the 6.4~keV line is observed to be narrow (FWHM $\lesssim$ a few thousand km~s$^{-1}$) and is commonly attributed to distant torus material (e.g., \citealt{2006MNRAS.368L..62N}). However, in some cases the line width suggests an origin in the outer BLR, as in NGC~7213 where the Fe K$\alpha$ FWHM ($\sim$2400 km~s$^{-1}$) matched that of broad H$\alpha$ \citep{2008MNRAS.389L..52B}. XRISM observations of NGC~4151 similarly reveal a broad component \citep{2024ApJ...973L..25X} that has a very similar profile to near simultaneous optical observations of H$\beta$ line (Noda et al. in prep). We perform a similar comparison here.

In NGC 1365, the broad component of the Fe~K$\alpha$ line has a width of FWHM $\sim 3060$~km~s$^{-1}$ ($\sigma \sim 1300$~km~s$^{-1}$), or when fitted with a {\tt diskline} model, constrained to a radius of $\sim 10^{4}~R_g$ (Section~\ref{subsec:Fe}), reminiscent of the BLR. Fig.~\ref{fig:opticalNIR}-{\em top} shows a comparison to the optical H$\beta$ line taken in 2013 \citep{2022ApJS..261....2K}. The H$\beta$ is narrower than the X-ray Fe~K line, but interestingly, it shows an asymmetric profile with an extended blue wing, similar to the X-ray line. In Fig.~\ref{fig:opticalNIR}-{\em bottom}, we compare to a NIR SINFONI spectrum from 2014 \citep{2019yCat..36220128F}, zooming in on the Pa$\alpha$. While lower signal-to-noise ratio, the Pa$\alpha$ appears much broader than the H$\beta$, which can be explained by extinction of the broad component of the H$\beta$ line. We find that the X-ray broad line width is very similar to the NIR broad line. Unlike Centaurus A for example, where a broad Fe K$\alpha$ component was detected but only a narrow component in the optical and NIR \citep{2025arXiv250702195B}, NGC~1365’s broad Fe K$\alpha$ is mirrored in width in the broad Pa$\alpha$, and in asymmetry in the optical H$\beta$ suggesting a common origin of the X-ray and optical/IR broad-line emission at BLR scales. 

It is interesting to note that both the optical and NIR spectra were taken in the Compton-thin/Seyfert 1 phase (\citealt{2019yCat..36220128F}; \citealt{2023MNRAS.518.2938T}; \citealt{2025A&A...693A..35J}), yet even in the Seyfert 2 XRISM observation, we detect a broad line with similar width to that seen in the NIR. Near-infrared, X-ray, and optical emission lines can be used to infer the distribution of gas and dust in the circumnuclear environment of AGN through more detailed radiative transfer modeling.

As described in Section \ref{subsec:Fe}, we model the broad component of the neutral Fe K$\alpha$ line using an intrinsic H97 profile convolved with a relativistic disk broadening component ({\tt diskline}). In this configuration, we find no evidence for variability in the broad line normalization (or flux) across the two observations. The H97 profile incorporates a fixed 2:1 flux ratio between the K$\alpha_1$ and K$\alpha_2$ lines, as expected from atomic physics \citep{1997PhRvA..56.4554H}. However, when replacing the H97 profiles with two narrow Gaussian components and one broad Gaussian component (an initial attempt to include two broad Gaussians for K$\alpha_1$ and K$\alpha_2$ proved degenerate, with one being suppressed), and allowing the narrow K$\alpha_1$ and K$\alpha_2$ normalizations to vary independently, we find that the K$\alpha_2$ normalization effectively vanishes in  Obs2 with $\Delta C = 6.6$ for one additional d.o.f (2.5$\sigma$). This deviation from the expected line ratio might be consistent with the changing ratio of  K$\alpha_1$ and K$\alpha_2$ in slightly ionized gas (for example, see Fig. 11 in \citealt{2004ApJS..155..675K}). 

The combination of a shifted Fe K edge, the anomalous ratio of  K$\alpha_1$ and K$\alpha_2$, and  the shape of the Fe K$\alpha$ complex are all consistent with the hypothesis that the Fe K$\alpha$ complex  with an effective velocity of  3,000 km~s$^{-1}$ FWHM arises in the slightly ionized gas of temperature 10,000-20,000 K of the broad line region (e.g., \citealt{2006LNP...693...77P}) rather than in the cold gas of a torus. 

Additionally, under this Gaussian modeling (allowing for deviations in the narrow K$\alpha_1$ and K$\alpha_2$ line ratios), we detect significant variability in the broad component, with Obs2 favoring a stronger contribution ($\Delta C = 22.7$ for two d.o.f; 4.4$\sigma$) of the broad component. If this variability is real, the timescale between the two observations ($\sim$5 months) implies an origin at $<5 \times 10^5 R_g$, consistent with the spatial scales inferred from the inner and outer radii of the disk broadening model discussed in Section~\ref{subsec:Fe}, and moreover, with the location of the ionized wind (Section~\ref{sec:origin-lines}). 

\subsection{Origin of the Ionized Emission Lines} 

\noindent NGC 1365 is a starburst galaxy, a class known to exhibit collisionally ionized emission lines, including ionized iron features as seen in M82 (e.g., \citealt{2023A&A...674A..77I}). However, NGC 1365 also hosts a powerful AGN capable of producing photoionized iron lines. These lines could arise from a photoionized wind exhibiting a classic P Cygni profile, stellar processes within the starburst, or shocks driven by the outflowing wind interacting with the interstellar medium, resulting in collisionally ionized emission and absorption.

Based on current Resolve data alone, it is not possible to unambiguously distinguish between these scenarios. However, we find some evidence supporting the interpretation that these emission lines originate from the outflow. In Section~\ref{subsec:pion} we find that in addition to the Fe\,\textsc{xxv} emission line, there is a tentative detection also of Fe\,\textsc{xxvi}, which implies a temperature (assuming collisional ionization) that is much hotter than typical $10^{6}-10^{7}$~K starburst systems (e.g, \citealt{2005ApJ...628..187G}). Previous studies of the soft X-ray emission below 2~keV with RGS data have revealed contributions from both photoionized and collisionally ionized plasma, with characteristic temperatures of $\sim$0.3 and $\sim$0.7~keV \citep{2009A&A...505..589G}, more consistent with expectations for star-forming regions.

Moreover, we also search for variability in the Fe\,\textsc{xxv} emission lines between the first and second Resolve observations that, if seen, would rule out a star-forming origin of the lines. We test the variability of the Fe\,\textsc{xxv} transitions using a phenomenological model of Gaussian lines. We subtract the two observations (Obs1 \& Obs2) to obtain a difference spectrum, which we model using four Gaussian components fixed at the redshifted transition energies of the Fe\,\textsc{xxv} K$\alpha$ complex (w, x, y, z; \citealt{NIST_ASD}). To account for any velocity offsets in the emission lines, we include a velocity shift model, and model the underlying continuum with a power-law. The inclusion of the Gaussian components improves the fit of this difference spectrum at the 2--3$\sigma$ significance level. Such variability suggests an origin in a dynamic photoionized gas, though higher signal-to-noise data are required to confirm this result.
An alternative test is the implied size of the hot gas emitting region. Assuming a typical broad-line region (BLR) density in the range $n_{\rm BLR} \sim 10^9\text{--}10^{11}~\mathrm{cm}^{-3}$, and a BLR temperature of $T_{ \rm BLR}\sim$~$2\times10^{4}$ K (e.g., \citealt{2006agna.book.....O}, \citealt{2016MNRAS.462.3570S}), we can infer the density of the hot phase assuming pressure equilibrium, $n_{\rm H} \sim \frac{n_{\rm BLR} \times T_{BLR}}{T_{hot}}$ where $T_{hot}\sim$7~keV from Table~\ref{tab:cie}, then using the observed emission measure of $\sim 3 \times 10^{63}~\mathrm{cm}^{-3}$ (Table~\ref{tab:cie}), we estimate the size of a uniform, spherical emitter from the relation $\mathrm{EM} = n_{\rm H} n_e V$, where $n_e = 1.2\,n_{\rm H}$. Solving for the radius using $V = \frac{4}{3}\pi R^3$, we obtain a radius in the range $R \sim (9.9 \times 10^{15}\text{--}2.1 \times 10^{17})~\mathrm{cm} \approx (3.2 \times 10^{-3}\text{--}6.9 \times 10^{-2})~\mathrm{pc} \approx (4\text{--}80)~\mathrm{light\text{-}days} \approx (1500\text{--}325000)~R_{\rm g}$. This range is consistent with an emission region at typical BLR distances, in line with the inferred distance to the wind and the size of broad Fe~K$\alpha$ component. This estimate depends on the assumed densities, and adopting lower densities (e.g., ISM-like values) would yield significantly larger emitting sizes. Additionally, the 5-month separation between the two observations would imply a characteristic size of the emitting region of $< 4 \times 10^{17}$ cm (0.13~pc), or approximately $< 6 \times 10^5\,R_g$ which is consistent with the wind location derived in Section \ref{sec:origin-lines} and is significantly more compact than the starburst ring lying at kiloparsec-scale distances.

Photoionized soft X-ray emission lines have previously been detected in NGC 1365, even in the Compton-thick state (e.g., \citealt{2015MNRAS.453.2558N}), and in one case, a potential P Cygni profile was reported \citet{2014ApJ...795...87B} in the Mg XII Ly$\alpha$ line. We find the ionization of the observed emission lines to be consistent with one of the photoionization phases identified by \citet{2015MNRAS.453.2558N} in high-resolution Chandra data (mainly complexes of O, Ne, Mg, and Si below 4 keV). These historical soft X-ray detections could represent lower ionization phases of the same photoionized gas responsible for the newly detected Fe\,\textsc{xxv} emission. To our knowledge, Fe\,\textsc{xxv} emission lines have not been previously reported in NGC 1365 in any state, but are now detectable thanks to XRISM's improved spectral capabilities.

Future variability studies and simultaneous high-resolution RGS observations (in prep) will be essential for confirming the origin of these emission features. Additional XRISM Resolve observations in a less obscured state may reveal strong emission lines, allowing better constraints on the physical conditions and kinematics of the emitting gas.
\section{Summary and Conclusions} \label{sec:summary}

\noindent We presented results from XRISM/Resolve observations of NGC 1365, obtained in February and July 2024. The source is observed in a persistent Compton-thick state (Fig. \ref{fig:XMM}), characterized by both a decrease in intrinsic hard X-ray luminosity and significant line-of-sight obscuration (Fig. \ref{fig:lc}). We summarize our main findings below:

\begin{itemize}
\item The spectrum reveals prominent Fe~\textsc{xxv} and Fe~\textsc{xxvi} absorption lines tracing a known highly ionized outflow, along with corresponding Fe~\textsc{xxv} and Fe~\textsc{xxvi} emission lines detected for the first time in this source (Fig. \ref{fig:resolve}).
\item The ionized iron emission lines are consistent with photoionized re-emission from the outflowing wind, forming a classic P Cygni profile (Fig. \ref{fig:pion}). Alternative origins for the ionized iron emission lines, including collisionally ionized gas from the circumnuclear starburst ring or shocked material driven by the wind itself. Producing both Fe~\textsc{xxv} and Fe~\textsc{xxvi} requires plasma temperatures of $kT \sim 7$ keV, which are higher than typically found in star-forming regions (Fig.~\ref{fig:cie}).
\item The outflow velocity (of the wind) implies a minimum launch radius of $\sim 10^4 R_{\mathrm{g}}$, while ionization-based arguments place an upper limit at comparable scales, jointly locating the wind within the broad-line region (BLR) (Section~\ref{sec:origin-lines}).
\item The Fe K$\alpha$ emission line is resolved and broadened ($\sigma \sim 1300$ km s$^{-1}$), with a width comparable to archival Pa$\alpha$, supporting a common origin of the X-ray and NIR gas flows at BLR-scales (Fig. \ref{fig:opticalNIR}).
\item Variability in the broad Fe K$\alpha$ component and disk broadening models of the Fe K$\alpha$ line (Section~\ref{subsec:Fe}) point to an origin at BLR scales, consistent with the inferred wind location.
\item Both the Fe K$\alpha$ and H$\beta$ lines show extended blue wings, suggesting an outflowing component, and supporting a shared origin between the wind and the broad Fe K emission within a dynamic BLR (Fig.~\ref{fig:Fe}).
\item It is noteworthy that many XRISM targets, such as NGC~4151 \citep{2024ApJ...973L..25X} and NGC~3783 \citep{2025A&A...699A.228M}, exhibit either obscurer absorption features from He-like and H-like iron, or emission only features from these same transitions, as seen in M81 \citep{2025ApJ...985L..41M}, NGC 7213 \citep{2025ApJ...994L..13K}, and Circinus~X-1 \citep{2025PASJ..tmp...30T}. In contrast, NGC~1365 displays both absorption and emission, which may indicate a higher covering fraction of the ionized gas or line-of-sight effects unique to this source.

\end{itemize}

The XRISM observations of NGC~1365, even in its unprecedented extended low-flux state, reveal a wealth of emission and absorption lines that are beginning to help us understand the complex relationship between inflows and outflows in highly variable, Changing-look AGN. Future Resolve observations in its higher flux states will offer an important point of comparison, helping to decipher how accretion drives (sometimes obscuring) outflows in these extreme systems.

\begin{acknowledgments}
FZ and EK acknowledge support by NASA Grant 80NSSC25K7065. PK acknowledges support from the NASA Hubble Fellowship grant HST-HF2-51534.001-A awarded by the Space Telescope Science Institute, which is operated by the Association of Universities for Research in Astronomy, Incorporated, under NASA contract NAS5-26555. EB and AJ acknowledges support from NASA Grant 80NSSC25K0082. ME acknowledges support from the U.S. Department of Energy by Lawrence Livermore National Laboratory under Contract DE-AC52-07NA27344. SO acknowledges support from the Japan Society for the Promotion of Science (JSPS) KAKENHI grant number 24K17104 (S.O.). Part of this research employs a list of Chandra datasets, obtained by the Chandra X-ray Observatory, contained in~\dataset[DOI:10.25574/cdc.509]{https://doi.org/10.25574/cdc.509}.

\end{acknowledgments}

\software{HEAsoft \citep{2014ascl.soft08004N},
    SPEX \citep{1996uxsa.conf..411K},
    SAS \citep{2004ASPC..314..759G}, 
    CIAO \citep{2006SPIE.6270E..1VF}.
}

\bibliography{bibliography}{}
\bibliographystyle{aasjournalv7}

\end{document}